\documentclass[aps,prb,twocolumn,showpacs,floatfix]{revtex4}

\usepackage{graphicx}
\usepackage{dcolumn}
\usepackage[english]{babel}

\begin{document}

\title{The Smeagol method for spin- and molecular-electronics}
\author{A. R. Rocha$^1$, V. M. Garc\'{\i}a-Su\'arez$^2$, S. Bailey$^3$,
C. Lambert$^3$, J. Ferrer$^2$  and S. Sanvito$^1$}
\email{sanvitos@tcd.ie}
\affiliation{$^1$ School of Physics, Trinity College, Dublin 2, IRELAND}
\affiliation{$^2$ Departamento de F\'{\i}sica, Universidad de Oviedo, 33007 Oviedo, SPAIN}
\affiliation{$^3$ Department of Physics, Lancaster University, Lancaster, UK}
\date{\today}

\begin{abstract}
{\it Ab initio} computational methods for electronic transport in
nanoscaled systems are an invaluable tool for the design of
quantum devices. We have developed a flexible and efficient
algorithm for evaluating $I$-$V$ characteristics of atomic
junctions, which integrates the non-equilibrium Green's function
method with density functional theory. This is currently
implemented in the package {\it Smeagol}. The heart of {\it
Smeagol} is our novel scheme for constructing the surface Green's
functions describing the current/voltage probes. It consists of a
direct summation of both open and closed scattering channels
together with a regularization procedure of the Hamiltonian, and
provides great improvements over standard recursive methods. In
particular it allows us to tackle material systems with
complicated electronic structures, such as magnetic transition
metals. Here we present a detailed description of {\it Smeagol}
together with an extensive range of applications relevant for the
two burgeoning fields of spin and molecular-electronics.
\end{abstract}

\pacs{75.47.Jn, 72.10.Bg, 73.63.-b}

\maketitle

\section{Introduction}\label{intro}
The study of electronic transport through devices comprising only
a handful of atoms is becoming one of the most fascinating branch
of modern solid state physics. The field was initiated with the
advent of the scanning tunnelling microscope (STM) \cite{STM} and
at present comprises a multitude of applications which span over
several disciplines and encompass different technologies, from
building blocks for revolutionary computer architectures, to
disposable electronics, to diagnostic tools for genetically driven
medicine. Clearly many of these devices will soon change and
enhance the quality of our daily life.

Fully functional molecular memories \cite{memory}
and logic gates \cite{gates1,gates2} have been already
demonstrated suggesting a possible roadmap to the post-silicon
era. These should produce future generations of computers,
together with magnetic data storage devices exceeding the
Terabit/in$^2$ storage limit. The readout of such high-density
data storage media will be achieved using nanoscale devices with
magnetic atomic point contacts \cite{PC1,PC2}.

At the same time hybrid molecular devices are becoming
increasingly popular in multifunctional sensor design,
demonstrating a sensitivity orders of magnitude superior to that
achievable with conventional methods. These molecular devices
include for example carbon nanotubes detectors for NO$_2$
\cite{NO2} and nerve agents \cite{nerve}, nanowire-based virus
detectors \cite{virus} and chemical sensors \cite{chem}. The near
future should see the development of on-chip nanolabs able to
sense a particular signature of gene or protein expression and
therefore be able to diagnose various diseases. These will be
formidable tools for the study of biological systems and in the
field of preventive medicine \cite{med}.

In addition to this large experimental activity an equally large
effort has been devoted to the development of efficient
computational methods for evaluating $I$-$V$ characteristics of
nanoscale devices. This is quite a remarkable theoretical
challenge since advanced quantum transport algorithms must be
combined with state of the art electronic structure methods.
Ideally these tools should be able to include strong correlation
as well as inelastic effects, and they should be suitable for
describing large systems (easily scalable methods). Furthermore in
order to compare directly to experiments the detailed knowledge of
the atomic configuration is needed.

Modern theory of quantum transport has developed a range of
methods for calculating transport in nanoscale conductors. Broadly
speaking these can be divided into two main classes: 1) steady
state algorithms, and 2) time dependent schemes. The first are
based upon the assumption that, regardless of the details of a
possible transient, a steady state is always achieved. The current
through the entire device is calculated as a balance of currents
entering and leaving a given scattering region, either using
scattering theory \cite{GF1,GF2,GF3,GF4,WF1,WF2,WF3}, or by
solving a master equation \cite{ME1,ME2,ME3}. A multitude of
variations over this generic scheme are available \cite{StefRev},
depending on the underlying assumption leading to the steady
state, the details of the electronic structure method employed,
and the way in which the external potential is introduced in the
calculation. Interestingly most of the methods can be demonstrated
to be applicable to cases of non-interacting electrons
\cite{stef1}, although their equivalence is not demonstrated for
the interacting case. Among these algorithms a particular place is
occupied by implementations of the non-equilibrium Green function
(NEGF) \cite{GF1,GF2,GF3,GF4} method within density functional
theory (DFT) \cite{DFT,KohnSham}. This approach, which is based on
equilibrium DFT to describe the electronic structures, has the
advantage of being conceptually simple, and computationally easy
and versatile to implement \cite{NEGF1,NEGF2,NEGF3,NEGF4,NEGF5}.

Time dependent methods are at an earlier stage of development.
These investigate the time-evolution of the electronic charge
density of a system, brought out of equilibrium by a
time-dependent perturbing potential. To the best of our knowledge
two fundamentally different methods have been proposed to date.
The first considers infinite non-periodic systems, with an
external potential introduced as a time dependent perturbation
\cite{stef1,cini}. The time-evolution of the density matrix is
studied with time-dependent density functional theory (TDDFT)
\cite{Gross,TDDFT}. An alternative approach consists of placing
the system of interest in a large capacitor. Such a capacitor is
charged at $t$=0 and the time-dependent de-charging process is
investigated \cite{Tchav,Max1}. Generally speaking these methods
are computationally intensive, since the need to perform the time
evolution adds to the computational overheads of standard static
schemes. However they should be able to address transport limits
such as Coulomb blockade or resonant tunnelling, which are
otherwise difficult to describe. Interestingly, important
information can be extracted from the static limit of the
time-dependent problem \cite{Max2,Kieron1}, and this can help in
designing more accurate static methods and in understanding their
limitations.

Here we present in details our recently developed quantum
transport code {\it Smeagol} \cite{Smeagol1}. {\it Smeagol} is a
DFT implementation of the non-equilibrium Green's function method,
which has been specifically designed for magnetic materials. The
main core of {\it Smeagol} is our original technique for
constructing the leads self-energies \cite{rgf}, which avoids the
standard problems of recursive methods \cite{GF1} and allows us to
describe devices having current/voltage probes with a complicated
electronic structure. In addition {\it Smeagol} has been
constructed to be a modular and scalable code, with particular
emphasis on heavy parallelization, to facilitate large scale
simulations. In its present form {\it Smeagol} is parallel over
$k$-space, real space and energy and furthermore it can deal with
spin-polarised systems, including spin non-collinearity. A partial
description of the code has already been provided \cite{Smeagol2},
which should be incorporated with this more detailed description.

The paper is organized as follows. In the next section we introduce our method and its main
technical implementations. In particular we set the problem, explain how to construct
the leads self-energies, and describe the strategy used for calculating
the electrostatic potential. Then we present a series of calculations for systems relevant
to either spin- or molecular-electronics. These address specific aspects of {\it Smeagol}
such as the electrostatics, the spin polarization and the spin non-collinearity. Finally
in the appendices we describe in more details the self-energy algorithm, we recall
the theoretical foundations of the NEGF formalism, and we
establish a connection with TDDFT.

\section{Non-equilibrium transport method}\label{theory}

In this section we describe in details our computational technique. The underlying
assumption used throughout this work is that all the quantities associated to the electronic
structure (Hamiltonian, density matrix, Green's functions, etc.) can be written over a
localized atomic orbital (LAO) basis set of some kind $\psi_{i\:\mu}=\psi_\mu(\vec{r}-\vec{R}_i)$,
where $\vec{R}_i$ is the position of the $i$-th nucleus and $\mu=n,l,m$  is a collective
index spanning, the angular momentum ($l,m$) and the orbital $n$.
Note that in general the index $n$ can run over different radial functions
corresponding to the same angular momentum, according to the multiple-zetas scheme
\cite{MZ}. In this way a generic operator $\hat{O}$ is represented by a finite
$N\times N$ matrix ($N$ is the total number of degrees of freedom in the system)
whose matrix elements are simply $O_{i\:\mu,~j\nu}$. Note also that the functions
$\psi_{i\:\mu}$ are generally non-orthogonal and the overlap matrix $S$ is defined
as $S_{i\:\mu,~j\nu}=\int \psi_\mu^*(\vec{r}-\vec{R}_i)\psi_\nu(\vec{r}-\vec{R}_j)\:
\mathrm{d}^3\vec{r}$.

\subsection{Problem setup}\label{negf}

\begin{figure}[ht]
\begin{center}
\includegraphics[width=7.0cm,clip=true]{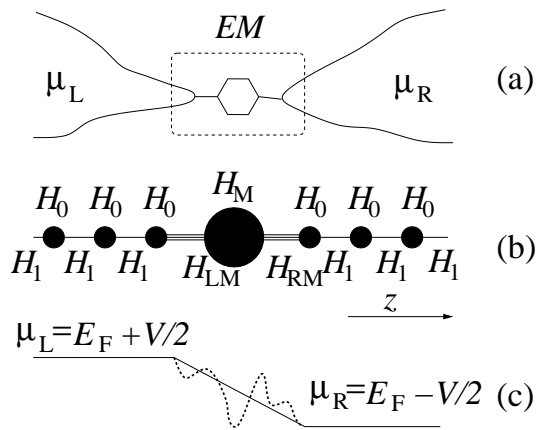}
\end{center}
\caption{\small{Schematic two terminal device. (a) Thermodynamical aspect: two leads
are kept respectively at the chemical potentials $\mu_\mathrm{L}$ and $\mu_\mathrm{R}$
and are able to exchange electrons with the central region (extended molecule {\it EM}).
(b) Hamiltonian description: two block diagonal infinite matrices describe the semi-infinite
current/voltage probes, and a finite matrix $H_\mathrm{M}$ describes the extended molecule.
$H_0$ is a finite Hamiltonian matrix describing one principal layer, while $H_1$ describes
the interaction between two adjacent principal layers. (c) Electrostatics: the two
current/voltage leads have a constant average potential $\mu_\mathrm{L/R}=E_\mathrm{F}\pm V/2$,
and the potential drop occurs within the extended molecule.}}
\label{Fig1}
\end{figure}

{\it Smeagol} has been designed to describe two-terminal
conductance experiments, where two current/voltage electrodes of
macroscopic size sandwich a nanometer-sized device (a molecule, an
atomic point contact, a tunneling barrier...). Let us present the
problem from three different perspectives: the thermodynamics, the
Hamiltonian and the electrostatics (see figure \ref{Fig1}).

From a thermodynamic point of view the system is modelled two bulk
leads and a central region. The latter includes the actual device
and, for reasons that will be clear later, part of the leads.
Therefore we call such a central region an ``extended molecule''
(EM). The two current/voltage leads are kept at two different
chemical potentials respectively $\mu_\mathrm{L}$ and
$\mu_\mathrm{R}$ and are able to exchange electrons with the EM.
Note that when the applied bias is zero
($\mu_\mathrm{L}=\mu_\mathrm{R}$), this system of interacting
electrons is in thermodynamic equilibrium and may be regarded as a
grand canonical ensemble. When the bias is applied however
$\mu_\mathrm{L}\ne\mu_\mathrm{R}$ and the current will flow. Then
the prescription for establishing the steady state is that of
adiabatically switching on the coupling between the leads and the
EM \cite{GF3,KEL1,KEL2}.

At the Hamiltonian level the system under investigation is
described by an infinite hermitian matrix ${\cal H}$. This however
has a rather regular structure. First notice that the two
semi-infinite current/voltage probes are defect-free crystalline
metals. These have a regular periodic structure and a unit cell
along which the direction of the transport can be defined. At this
point it is important to notice that because of the LAO basis set
this matrix will be rather sparse. It is convenient to introduce
the concept of principal layer (PL). A principal layer is the
smallest cell that repeats periodically in the direction of the
transport constructed in such a way to interact only with the
nearest neighbour PLs. This means that {\it all} the matrix
elements between atoms belonging to two non-adjacent PLs vanish.
For example take a linear chain of hydrogen atoms described by a
nearest neighbour tight-binding model then one atom forms the unit
cell. However if nearest and next nearest neighbour elements are
included then the PL will contain two atoms etc (for examples see
appendix \ref{H1problem}).

We then define $H_0$ as the $N\times N$ matrix describing all
interactions within a PL, where $N$ is the total number of degrees
of freedom (basis functions) in the PL (note that we use
calligraphic symbols - ${\cal H}$ - for infinitely-dimensional
matrices and capitalized letters - $H$ - for finite matrices).
Similarly $H_1$ is the $N\times N$ matrix describing the
interaction between two PLs. Finally $H_\mathrm{M}$ is the
$M\times M$ matrix describing the extended molecule and
$H_\mathrm{LM}$ ($H_\mathrm{RM}$) is the $N\times M$ matrix
containing the interaction between the last PL of the left-hand
side (right-hand side) lead and the extended molecule. The final
form of ${\cal H}$ is
\begin{equation}
{\cal H}
=
\left(\begin{array}{ccccccccccc}
.&.&.&.&.&.&.&.&.&.&\\
.&0&H_{-1}&H_0&{H}_1&0&.&.&.&.&.\\
.&.&0&H_{-1}&H_0&{H}_\mathrm{LM}&0&.&.&.&.\\
.&.&.&0& H_\mathrm{ML}& H_\mathrm{M} &H_\mathrm{MR}&0&.&.&.\\
.&.&.&.&0&H_\mathrm{RM}&H_0&H_1&0&.&.\\
.&.&.&.&.&0&H_{-1}&H_0&H_1&0&.\\
.&.&.&.&.&.&.&.&.&.&.\\
\end{array}\right)\;.
\end{equation}\label{Htot}
For a system which preserves time-reversal symmetry
$H_{-1}=H_1^\dagger$, $H_\mathrm{ML}=H_\mathrm{LM}^\dagger$ and
$H_\mathrm{MR}=H_\mathrm{RM}^\dagger$. In this form ${\cal H}$ has
the same structure as the Hamiltonian of a one-dimensional system
as shown in figure \ref{Fig1}b. However this is not the most
general situation and does not apply if a magnetic field is
present for example.

Note that the overlap matrix ${\cal S}$ has exactly the same
structure of ${\cal H}$. Therefore we adopt the notation $S_0$,
$S_1$, $S_\mathrm{LM}$, $S_\mathrm{RM}$ and $S_\mathrm{M}$ for the
various blocks of ${\cal S}$, in complete analogy with their
Hamiltonian counterparts. Here the principal layer, defined by
${\cal H}$ is used for both the ${\cal S}$ and the ${\cal H}$
matrix, even though the range of ${\cal S}$ can be considerably
shorter than that of ${\cal H}$.

Let us now discuss the electrostatics of the problem (figure
\ref{Fig1}c). The main consideration here is that the
current/voltage probes are made from good metals and therefore
preserve local charge neutrality. For this reason the effect of an
external bias voltage on the leads will produce a rigid shift of
the whole spectrum, i.e. of all the on-site energies. In contrast
a non-trivial potential profile will develop over the extended
molecule, which needs to be calculated self-consistently.
Importantly the resulting self-consistent electrostatic potential
must match that of the leads at the boundaries of the EM. If this
does not happen, the potential profile will develop a
discontinuity with the generation of spurious scattering.
Therefore, in order to achieve a good match of the electrostatic
potential, several layers of the leads are usually included in the
extended molecule. Their number ultimately depends upon the
screening length of the leads, but in most situations a few
(between two and four) atomic planes are sufficient.

Even in the case of extremely short screening length, it is
however good practice to include a few planes of the leads in the
extended molecule because the electrodes generally have
reconstructed surfaces, which might undergo additional geometrical
reconstructions when bonding to the nanoscale device (e.g.
molecules attached to metallic surfaces through corrosive chemical
groups).

We conclude this section with some comments about the application
of periodic boundary conditions in the transverse direction
perpendicular to that of the transport. The setup of a typical
experiment is that of two very large current voltage probes
sandwiching a tiny region which is responsible for most of the
resistance. The ideal description would be that of two infinite
leads (with infinite cross sections) and a finite scattering
region. Unfortunately this problem is intractable since both $H_0$
and $H_1$ become infinitely-dimensional. Therefore one has to
consider some approximations.

The first option is to use leads with finite cross section. In
this case, no periodic boundary conditions are required and the
whole system is quasi-one-dimensional. However special care must
be taken when choosing the cross section of the leads in order to
avoid quantum confinement effects. It is also worth noting that
leads with very small cross section make the use of the Landauer
formula for transport \cite{LBYP} questionable. As a rule of thumb
the linear dimension of the cross section should be several times
the Fermi wavelength of the material forming the leads and there
should be several open scattering channels.

The second option is to use periodic boundary conditions. In this
case the system is repeated periodically in the transverse
direction, meaning that the extended molecule is also repeated
periodically. Clearly quantum confinement effects are eliminated,
but one should be particularly careful in order to eliminate the
spurious interaction between the mirror images of the extended
molecule. Therefore large unit cells must be employed even when
periodic boundary conditions are used. However from a formal point
of view the use of periodic boundary conditions does not change
the problem setup. All the matrices ($H_0$, $H_1$ etc.) now depend
on the transverse $k$-vector used, and the infinite problem
transforms in a collection of $k$-dependent quasi one-dimensional
problems. This dependence is implicitly assumed whenever necessary
throughout the rest of the paper.

\subsection{Green's functions for an open system}\label{negfdft}

We are dealing with an infinite-dimensional hermitian problem,
which is intractable, because the wave-functions deep inside the
leads have a plane-wave form. These can be calculated by computing
the band-structure of an infinite chain of PL's. The main
computational effort is therefore focussed upon the problem of
describing the scattering of plane-waves from one lead to the
other across the EM. The problem is solved by computing the
retarded Green's function ${\cal G}^\mathrm{R}$ for the whole
system by solving the Green's function equation
\begin{equation}
[\epsilon^+{\cal S}-{\cal H}]\:{\cal G}^\mathrm{R}(E)={\cal I}\:,
\end{equation}
where ${\cal I}$ is an infinitely-dimensional identity matrix,
$\epsilon^+=\lim_{\delta\rightarrow0^+}E+i\delta$ and $E$ is the
energy. The same equation explicitly using the block-diagonal
structure of both the Hamiltonian and the overlap matrix (we drop
the symbol ``R'' indicating the retarded quantities) is of the
form
\begin{widetext}
\begin{equation}
\left(\begin{array}{ccc}
\epsilon^+{\cal S}_\mathrm{L}-{\cal H}_\mathrm{L}&\epsilon^+{\cal S}_\mathrm{LM}-{\cal H}_\mathrm{LM}&0\\
\epsilon^+{\cal S}_\mathrm{ML}-{\cal H}_\mathrm{ML}&\epsilon^+S_\mathrm{M}-H_\mathrm{M}&\epsilon^+{\cal S}_\mathrm{MR}-{\cal H}_\mathrm{MR}\\
0& \epsilon^+{\cal S}_\mathrm{RM}-{\cal H}_\mathrm{RM}&\epsilon^+{\cal S}_\mathrm{R}-{\cal H}_\mathrm{R}\\
\end{array}\right)
\left(\begin{array}{ccc}
{\cal G}_\mathrm{L}&{\cal G}_\mathrm{LM}&{\cal G}_\mathrm{LR}\\
{\cal G}_\mathrm{ML}&G_\mathrm{M}&{\cal G}_\mathrm{MR}\\
{\cal G}_\mathrm{RL}&{\cal G}_\mathrm{RM}&{\cal G}_\mathrm{R}\\
\end{array}\right)=
\left(\begin{array}{ccc}
{\cal I}&0&0\\
0&I_\mathrm{M}&0\\
0&0&{\cal I}\\
\end{array}\right)\;,
\label{Eq3}
\end{equation}
\end{widetext}
where we have partitioned the Green's functions ${\cal G}$ into
the infinite blocks describing the left- and right-hand side leads
${\cal G}_\mathrm{L}$ and ${\cal G}_\mathrm{R}$, those describing
the interaction between the leads and extended molecule ${\cal
G}_\mathrm{LM}$, ${\cal G}_\mathrm{RM}$, the direct scattering
between the leads ${\cal G}_\mathrm{LR}$, and the finite block
describing the extended molecule $G_\mathrm{M}$. We have also
introduced the matrices ${\cal H}_\mathrm{L}$, ${\cal
H}_\mathrm{R}$, ${\cal H}_\mathrm{LM}$, ${\cal H}_\mathrm{RM}$ and
their corresponding overlap matrix blocks, indicating respectively
the left- and right-hand side leads Hamiltonian and the coupling
matrix between the leads and the extended molecule.
$H_\mathrm{M}$ is an $M\times M$ matrix and $I_\mathrm{M}$ is the
$M\times M$ unit matrix. The infinite matrices, ${\cal
H}_\mathrm{L}$ and ${\cal H}_\mathrm{R}$ describe the leads and
have the following block-diagonal form
\begin{equation}
{\cal H}_\mathrm{L}=
\left(\begin{array}{ccccc}
\ddots&\ddots&\ddots&\ddots&\vdots\\
0&H_{-1}&H_0&H_1&0\\
\dots&0&H_{-1}&H_0&H_1\\
\dots&\dots&0&H_{-1}&H_0\\
\end{array}\right)\:,
\end{equation}
with similar expressions for ${\cal H}_\mathrm{R}$ and the overlap
${\cal S}$ matrix counterparts. In contrast the coupling matrices
between the leads and the extended molecule are infinite
dimensional matrices whose elements are all zero except for a
rectangular block coupling the last PL of the leads and the
extended molecule. For example we have
\begin{equation}
{\cal H}_\mathrm{LM}=
\left(\begin{array}{c}
\vdots \\
0\\
H_\mathrm{LM}\\
\end{array}\right)\:.
\end{equation}

The crucial step in solving equation (\ref{Eq3}) is to write down
the corresponding equation for the Green's function involving the
EM and surface PL's of the left and right leads and then evaluate
the retarded Green's function for the extended molecule
$G_\mathrm{M}^\mathrm{R}$. This has the form
\begin{equation}\label{gretarded}
G_\mathrm{M}^\mathrm{R}(E)=\left[\epsilon^+S_\mathrm{M}-H_\mathrm{M}-
\Sigma_\mathrm{L}^\mathrm{R}(E)-\Sigma_\mathrm{R}^\mathrm{R}(E)\right]^{-1}\:,
\label{rgfem}
\end{equation}
where we have introduced the retarded self-energies for the left- and right-hand side lead
\begin{equation}
\Sigma_\mathrm{L}^\mathrm{R}(E)=(\epsilon^+S_\mathrm{ML}-H_\mathrm{ML})
G_\mathrm{L}^{0\mathrm{R}}(E)\:
(\epsilon^+S_\mathrm{LM}-H_\mathrm{LM})\: \label{senleft}
\end{equation}
and
\begin{equation}
\Sigma_\mathrm{R}^\mathrm{R}(E)=(\epsilon^+S_\mathrm{MR}-H_\mathrm{MR})
G_\mathrm{R}^{0\mathrm{R}}(E)\:
(\epsilon^+S_\mathrm{RM}-H_\mathrm{RM})\:. \label{senright}
\end{equation}

Here $G_\mathrm{L}^{0\mathrm{R}}$ and
$G_\mathrm{R}^{0\mathrm{R}}$ are the retarded {\it surface}
Green's function of the leads, i.e. the leads retarded Green's
functions evaluated at the PL neighboring the extended molecule.
Formally $G_\mathrm{L}^{0\mathrm{R}}$ ($G_\mathrm{R}^{0\mathrm{R}}$)
corresponds to the right lower (left
higher) block of the retarded Green's function for the whole
left-hand side (right-hand side) lead. These are simply
\begin{equation}
{\cal G}_\mathrm{L}^{0\mathrm{R}}(E)=[\epsilon^+{\cal S}_\mathrm{L}-{\cal H}_\mathrm{L}]^{-1}\:
\label{Eq9}
\end{equation}
and
\begin{equation}
{\cal G}_\mathrm{R}^{0\mathrm{R}}(E)=[\epsilon^+{\cal S}_\mathrm{R}-{\cal H}_\mathrm{R}]^{-1}\:.
\label{Eq10}
\end{equation}
Note that ${\cal G}_\mathrm{L}^{0\mathrm{R}}$ (${\cal
G}_\mathrm{R}^{0\mathrm{R}}$) is not the same as ${\cal
G}_\mathrm{L}^{\mathrm{R}}$ (${\cal G}_\mathrm{R}^{\mathrm{R}}$)
defined in equation (\ref{Eq3}). In fact the former are the
Green's functions for the semi-infinite leads in isolation, while
the latter are the same quantities for the leads attached to the
scattering region. Importantly one does not need to solve the
equations (\ref{Eq9}) and (\ref{Eq10}) for calculating the leads
surface Green's functions and a closed form avoiding the inversion
of infinite matrices can be provided \cite{rgf}. This will be
discussed in what follows and in appendix \ref{H1problem}.

Let us conclude this section with a few comments on the results
obtained. The retarded Green's function $G_\mathrm{M}^\mathrm{R}$
contains all the information about the electronic structure of the
extended molecule attached to the leads. In its close form given
by the equation (\ref{rgfem}) it is simply the retarded Green's
function associated to the effective Hamiltonian matrix
$H_\mathrm{eff}$
\begin{equation}
H_\mathrm{eff}=H_\mathrm{M}+\Sigma_\mathrm{L}^\mathrm{R}(E)+\Sigma_\mathrm{R}^\mathrm{R}(E)\:.
\label{heff}
\end{equation}
Note that $H_\mathrm{eff}$ is not hermitian since the
self-energies are not hermitian matrices. This means the the
number of particles in the extended molecule is not conserved, as
expected by the presence of the leads. Moreover, since
$G_\mathrm{M}^\mathrm{R}$ contains all the information about the
electronic structure of the extended molecule in equilibrium with
the leads, it can be directly used for extracting the zero-bias
conductance $G$ of the system. In fact one can simply apply the
Fisher-Lee \cite{GF1,fisherlee} relation and obtain
\begin{equation}
G=\frac{2e^2}{h}\mathrm{Tr}[\Gamma_\mathrm{L}\:
G^\mathrm{R\dagger}_\mathrm{M}\:\Gamma_\mathrm{R}\:G^\mathrm{R}_\mathrm{M}]\:,
\label{zbcond}
\end{equation}
where
\begin{equation}\label{gamma}
\Gamma_\alpha(E)=i[\Sigma^\mathrm{R}_\alpha(E)-\Sigma^\mathrm{R}_\alpha(E)^\dagger]\:.
\end{equation}
In equation (\ref{zbcond}) all the quantities are evaluated at the
Fermi energy $E_\mathrm{F}$. Clearly
$\mathrm{Tr}[\Gamma_\mathrm{L}\:G^\mathrm{R\dagger}_\mathrm{M}\:\Gamma_\mathrm{R}\:
G^\mathrm{R}_\mathrm{M}](E)$ is simply the energy dependent total
transmission coefficient $T(E)$ of standard scattering theory
\cite{LBYP}.

Finally note that what we have elaborated so far is an alternative
way of solving a scattering problem. In standard scattering theory
one first computes the asymptotic current carrying states deep
into the leads (scattering channels) and then evaluate the quantum
mechanical probabilities for these channels to be reflected and
transmitted through the extended molecule \cite{LBYP}. In this
case the details of the scattering region are often reduced to a
matrix describing the effective coupling between the two surface
PLs of the leads \cite{rgf}. In contrast the use of (\ref{zbcond})
describes an alternative though equivalent approach, in which the
leads are projected out to yield a reduced matrix describing an
effective EM. The current through surface PL's perpendicular to
the transport direction are the same \cite{TCH1}, the two
approaches are equivalent and there is no clear advantage in using
either one or the other. However, when the Hamiltonian matrix of
the scattering region $H_\mathrm{M}$ is not known {\it a priori},
then the NEGF method offers a simple way of setting up a
self-consistent procedure.

\subsection{Steady-state and self-consistent procedure}\label{sfc}

Consider now the case in which the matrix elements of the
Hamiltonian of the system are not known explicitly, but only their
functional dependence upon the charge density $\rho$, ${\cal
H}={\cal H}[\rho]$, is known. This is the most common case in
standard mean field electronic structure theory, such as DFT. If
no external bias is applied to the device (linear response limit)
the Hamiltonian of the system can be simply obtained from a
standard equilibrium DFT calculation and the procedure described
in the previous section can be applied without any modification.
However, when an external bias $V$ is applied, the charge
distribution of the extended molecule will differ from that at
equilibrium since both the net charge and the electrical
polarization are affected by the bias. This will determine a new
electrostatic potential profile with different scattering
properties.

These modifications will affect only the extended molecule, since
our leads are required to preserve local charge neutrality. This
means that the charge density and therefore the Hamiltonian of the
leads are not modified by the external bias applied. As discussed
at the beginning the only effect of the external bias over the
current/voltage electrodes is that of a rigid shift of the on-site
energies. The Hamiltonian then takes the form
\begin{equation}
{\cal H}=
\left(\begin{array}{ccc}
{\cal H}_\mathrm{L}+{\cal I}\:eV/2&{\cal H}_\mathrm{LM}&0\\
{\cal H}_\mathrm{ML}&H_\mathrm{M}&{\cal H}_\mathrm{MR}\\
0&{\cal H}_\mathrm{RM}&{\cal H}_\mathrm{R}-{\cal I}\:eV/2\\
\end{array}\right)\:,
\label{hbias}
\end{equation}
Note that the coupling matrices between the leads and the extended
molecule are also not modified by the external bias, since by
construction the charge density in the surface planes of the
extended molecule matches exactly that of the leads.

The Hamiltonian of the extended molecule
\begin{equation}\label{Hamiltonian}
H_\mathrm{M}=H_\mathrm{M}[\rho]
\end{equation}
depends on the density matrix, which is calculated using
the lesser Green's function $G^<_\mathrm{M}$ \cite{GF1,GF2,GF3,GF4,KEL1,KEL2}
\begin{equation}\label{density}
\rho=\frac{1}{2\pi i}\int\mathrm{d}E\,G^<_\mathrm{M}(E)\;,
\end{equation}
so a procedure must be devised to compute this quantity.

In equilibrium,
$G^<(E)=-2i\mathrm{Im}\,\left[G^\mathrm{R}(E)\right]f(E-\mu)$, so
it is only necessary to consider the retarded Green's function,
given by equation (\ref{gretarded}). Alternatively,
$G^\mathrm{R}$ may be obtained from the eigenvectors of $\cal H$.

Out of equilibrium, however, the presence of the leads establishes
a non-equilibrium population in the extended molecule and $G^<$ is
no longer equal to $-2i\,\mathrm{Im}\,\left[G^\mathrm{R}\right]f(E-\mu)$.
The non-equilibrium
Green's function formalism \cite{GF1,GF2,GF3,GF4,KEL1,KEL2}
provides the correct expression (see appendix \ref{AppB}):
\begin{equation}\label{rgfem2}
G^<_{\mathrm M}(E)=iG_{\mathrm M}^{\mathrm R}(E)[\Gamma_{\mathrm
L}f(E-\mu_{\mathrm L})+\Gamma_{\mathrm R} f(E-\mu_{\mathrm
R})]G_{\mathrm M}^{\mathrm R\dag}(E)
\end{equation}
where $\mu_\mathrm{L/R}=\mu\pm eV/2$, $f(x)$ is the Fermi function for a given
temperature $T$,
\begin{equation}
\Gamma_\mathrm{L/R}=\Gamma_\mathrm{L/R}(E\pm eV/2)
\label{GammaL}
\end{equation}
and $G_\mathrm{M}^\mathrm{R}(E)$ is given again by the equation (\ref{rgfem}) where
now we replace $\Sigma_\mathrm{L/R}(E)$ with
$\Sigma_\mathrm{L/R}=\Sigma_\mathrm{L/R}(E\pm eV/2)$.

Finally the self consistent procedure is as follows. 
First a trial charge density $\rho^0$
is used to compute $H_\mathrm{M}$ from equation
(\ref{Hamiltonian}). Then $\Gamma_\mathrm{L}$, $\Gamma_\mathrm{R}$
and $G_\mathrm{M}^\mathrm{R}$ are calculated from equations
(\ref{GammaL}), and (\ref{rgfem}). These quantities
are used to compute the $G^<_\mathrm{M}$ of equation (\ref{rgfem2}),
which is fed back in equation (\ref{density}) to find a new
density $\rho^1$. This process is iterated until a self-consistent solution
is achieved, which is when
$||\rho^{j}-\rho^{j+1}||\ll1$. At this point
the problem is identical to that solved in the previous section
(since the whole ${\cal H}$ is now determined) and the current $I$
can be calculated using \cite{meir}
\begin{equation}
I=\frac{e}{h}\int\mathrm{d}E \ \mathrm{Tr}[\Gamma_\mathrm{L}\:
G^\mathrm{R\dagger}_\mathrm{M}\:\Gamma_\mathrm{R}\:G^\mathrm{R}_\mathrm{M}]
\left[f\left(E-\mu_\mathrm{L}\right) -
f\left(E-\mu_\mathrm{R}\right)\right]\:. \label{current}
\end{equation}
Note that now the transmission coefficient depends on both the
energy $E$ and the bias $V$.

Let us conclude this section with a note on how to perform the
integrals of equations (\ref{density}) and (\ref{current}). The
one for the current is trivial since the two Fermi functions
effectively cut the integration to give a narrow energy window
between the chemical potentials of the leads. In addition the
transmission coefficient, with the exception of some tunneling
situations, is usually a smooth function of the energy.

In contrast the integration leading to the density matrix
(\ref{density}) is more difficult, since the integral is unbound
and the Green's function has poles over the real energy axis. This
however can be drastically simplified by adding and subtracting
the term
$G_\mathrm{M}^\mathrm{R}\Gamma_\mathrm{R}G_\mathrm{M}^\mathrm{R\dagger}
f\left(E-\mu_\mathrm{L}\right)$ and by re-writing the integral
(\ref{density}) as the sum of two contributions
$\rho=\rho_\mathrm{eq}+\rho_V$
\begin{equation}
\rho_\mathrm{eq}=-\frac{1}{\pi}\int \mathrm{d}E \ \mathrm{Im}
\left[G_\mathrm{M}^\mathrm{R}\right]
f\left(E-\mu_\mathrm{L}\right), \label{D_equil}
\end{equation}
and
\begin{equation}
\rho_V= \frac{1}{2\pi}\int \mathrm{d}E \ G_\mathrm{M}^\mathrm{R}
{\Gamma}_\mathrm{R} G_\mathrm{M}^\mathrm{R\dagger}
\left[ f\left(E-\mu_\mathrm{R}\right) -
f\left(E-\mu_\mathrm{L}\right)\right].
\label{D_out_equil}
\end{equation}

$\rho_\mathrm{eq}$ can be interpreted as the density matrix at
equilibrium, {\it i.e.} the one obtained when both the reservoirs
have the same chemical potential $\mu_\mathrm{L}$, while $\rho_V$
contains all the corrections due to the non-equilibrium
conditions. Computationally $\rho_V$ is bound by the two Fermi
functions of the leads, as for the current $I$, and therefore one
needs to perform the integration only in the energy range between
the two chemical potentials. In contrast $\rho_\mathrm{eq}$ is
unbound, but the integral can be performed in the complex plane
using a standard contour integral technique \cite{lang}, since
$G_\mathrm{M}^\mathrm{R}$ is both analytical and smooth.

\subsection{Surface Green's functions}\label{leads}

Let us now return to the question of how to calculate the
self-energies for the leads.
From the equations (\ref{senleft}) and (\ref{senright}) it is clear that the problem is
reduced to that of computing the retarded surface Green functions
for the left- ($G_\mathrm{L}^\mathrm{0R}$) and right-hand side
($G_\mathrm{R}^\mathrm{0R}$) lead respectively. This does not
require any self-consistent procedure since the Hamiltonian is
known and it is equal to that of the bulk leads plus a rigid shift
of the on-site energies. However the calculation should be
repeated several times since the $\Sigma$'s depend both on the
energy and the $k$-vector. Therefore it is crucial to have a
stable algorithm.

There are a number of techniques in the literature to calculate
the surface Green's functions of a semi-infinite system. These
range from recursive methods \cite{GF1,nardelli} to
semi-analytical constructions \cite{rgf}. In {\it Smeagol} we have
generalized the scheme introduced by Sanvito {\it et. al.}
\cite{rgf} to non-orthogonal basis sets. This method gives us a
prescription for calculating the retarded surface Green's function
exactly. The main idea is to construct the Green's function for an
infinite system as a summation of Bloch states with both real and
imaginary wave-vectors, and then to apply the appropriate boundary
conditions to obtain the Green's function for a semi-infinite
lead.
\begin{figure}[h]
\begin{center}
\includegraphics[width=7.5cm,clip=true]{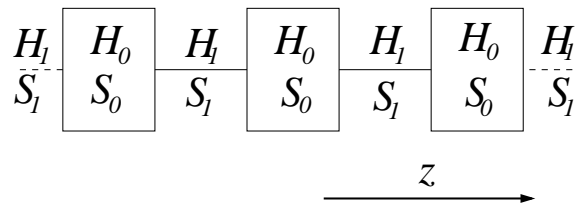}
\end{center}
\caption{Infinite periodic system used as current/voltage probe
and schematic diagram of the Hamiltonian. $H_0$ and $S_0$ are the
matrices describing the Hamiltonian and the overlap within a PL,
while $H_1$ and $S_1$ are the same quantities calculated between
two adjacent PLs. The arrow indicates the direction of transport.}
\label{Fig2}
\end{figure}

As explained above the Hamiltonian and the overlap matrices are
arranged in a tridiagonal block form, having respectively $H_0$
and $S_0$ on the diagonal, and $H_1$ and $S_1$ as the first off
diagonal blocks (see figure \ref{Fig2}). Since we are dealing with
an infinite periodic quasi one-dimensional system, the
Schr\"odinger equation can be solved for Bloch states
\begin{equation}
\psi_z=n_{k}^{1/2}e^{ikz}\phi_{k} \label{bloch}
\end{equation}
and reads
\begin{equation}\label{schbloch}
\left[K_0+K_1e^{ik}+
K_{-1}e^{-ik}\right]\phi_k=0\;,
\end{equation}
where $z=a_0j$ with $j$ integer and $a_0$ the separation between
principal layers, $k$ is the wave-vector along the direction of
transport (in units of $\pi/a_0$), $\phi_k$ is a $N$-dimensional
column vector and $n_{k}$ a normalization factor. Here we
introduce the $N\times N$ matrices
\begin{eqnarray}
K_0=H_{0}-E S_{0}\:, \\
K_1=H_{1}-E S_{1}\:, \\
K_{-1}=H_{-1}-E S_{-1}\:.
\end{eqnarray}

Since the Green's functions are constructed at a given energy our
task is to compute $k(E)$ (both real and complex) instead of
$E(k)$ as conventionally done in band theory. A numerically
efficient method to solve the ``inverse'' secular equation
$k=k(E)$ is to map it onto an equivalent eigenvalue problem. It is
simple to demonstrate \cite{rgf} that the eigenvalues of the
following $2N\times 2N$ matrix
\begin{equation}
\left(
\begin{array}{cc}
-{K_{1}}^{-1}{K_{0}} & -{K_{1}}^{-1}K_{-1} \\
I_\mathrm{N} & 0
\end{array}
\right)\label{secular}
\end{equation}
are $e^{ik}$ and that the upper $N$ component of the eigenvectors
are the vectors $\phi_k$. Clearly for the solution of this
eigenvalue problem one needs to invert $K_1$. However, since $K_1$
is determined by the details of the physical system, the choice of
basis set and of principal layer may be singular or severely
ill-conditioned. This problem often originates from the fact that
a few states within a PL do not couple to states in the
nearest-neighbouring PLs, but it can also be due to the symmetry
of the problem. For example in the case of {\it ab initio} derived
matrices this becomes unavoidable when one considers transition
metals, where the strongly localized $d$ shells coexist with
rather delocalised $s$ electrons. A possible solution to this
problem is to consider an equivalent {\it generalized} eigenvalue
problem, which does not require matrix inversion. However this
solution is not satisfactory for two reasons. First the matrices
still remain ill-conditioned and the general algorithm is rather
unstable. Secondly for extreme cases we have discovered that the
generalized eigenvalue solver cannot return meaningful eigenvalues
(divisions by zero are encountered). We therefore decide to use an
alternative approach constructing a regularization procedure for
eliminating the singularities of $K_1$. This must be performed
before starting the actual calculation of the Green's functions.
We will return on this aspect in appendix \ref{H1problem}. For the
moment we assume that $K_1$ has been regularized and it is neither
singular nor ill-conditioned.

When using orthogonal basis sets the knowledge of $k$ and
$\{\phi_{k}\}$ is sufficient to construct the retarded Green's
function for the doubly-infinite system, which has the form
\cite{rgf}
\begin{equation}\label{leadstotalG}
G_{zz^\prime}=\left\{
\begin{array}{cc}
\sum^N_l\phi_{k_l}e^{i k_l \left(z-z^\prime\right)}\tilde{\phi}_{k_l}^\dagger{V}^{-1} & z \geq z^\prime\\
\sum^N_l\phi_{\bar{k}_l}e^{i \bar{k}_l \left(z-z^\prime\right)}
\tilde{\phi}_{\bar{k}_l}^\dagger{V}^{-1} & z \leq z^\prime
\end{array}\right. \;,
\label{gfans}
\end{equation}
where the summation runs over both real and imaginary $k_l$. In
equation (\ref{gfans}) $k_l$ ($\bar{k}_l$) are chosen to be the
right-moving or right-decaying (left-moving or left-decaying)
Bloch states, i.e. those with either positive group velocity or
having $k$-vector with positive imaginary part (negative group
velocity or negative imaginary part). $\{\phi_{k_l}\}$ are the
corresponding vectors, and ${V}$ is defined in reference
[\onlinecite{rgf}]. Finally $\{\tilde{\phi}_{k_l}\}$ is just the
dual of $\{\phi_{k_l}\}$ obtained from
\begin{eqnarray}
\tilde{\phi}_{k_l}^\dagger\phi_{k_m}=\delta_{lm}\\
\tilde{\phi}_{\bar{k}_l}^\dagger\phi_{\bar{k}_m}=\delta_{lm}\\
\end{eqnarray}

In the case of a non-orthogonal basis set the same expression is
still valid if ${V}$ is now defined as follows
\begin{equation}
{V}=\sum^N_l\left(H^\dagger_1-E S^\dagger_1\right)\left[\phi_{k_l}
e^{-i k_l}\phi_{k_l}^\dagger-\phi_{\bar{k}_l}e^{-i \bar{k}_l}
\phi_{\bar{k}_l}^\dagger\right].
\end{equation}

Finally the surface Green's functions for a semi-infinite system
can be obtained from those of the doubly-infinite one by an
appropriate choice of boundary conditions. For instance if we
subtract the term
\begin{equation}
\Delta_{z}\left(z^\prime-z_0\right)=\sum^N_{l,h}\phi_{\bar{k}_h}e^{i \bar{k}_h
\left(z-z_0\right)}\phi_{\bar{k}_h}^\dagger\phi_{k_l}e^{i k_l\left(z_0-z^\prime\right)}
\phi_{k_l}^\dagger{V}^{-1}\;,
\end{equation}
from $G_{zz^\prime}$ of equation (\ref{gfans}) we obtain a new retarded Green's
function vanishing at $z=z_0$. Note that $\Delta_{z}\left(z^\prime-z_0\right)$ is a linear
combination of eigenvectors and therefore does not alter the causality of
$G$.

In this way we obtain the final expression for the retarded surface Green's
functions of both the left- and right-hand side lead
\begin{eqnarray}
{G}_\mathrm{L}^{0\mathrm{R}} & = & \left[ I_\mathrm{N} - \sum_{l,h} \phi_{\bar{k}_h}
e^{-i \bar{k}_h}\phi_{\bar{k}_h}^\dagger \phi_{k_l}e^{i k_l}\phi_{k_l}^\dagger\right]
{V}^{-1}, \\
{G}_\mathrm{R}^{0\mathrm{R}} & = & \left[ I_\mathrm{N} - \sum_{l,h} \phi_{k_h}e^{i k_h}\phi_{k_h}^\dagger
\phi_{\bar{k}_l}e^{-i \bar{k}_l}\phi_{\bar{k}_l}^\dagger\right]{V}^{-1}.
\end{eqnarray}
These need to be computed at the beginning of the calculation only.

\subsection{DFT implementation and electrostatics}\label{dft}

The formalism presented in sections \ref{negf} through \ref{leads} is rather
general and is not specific of a particular functional dependence of the Hamiltonian
upon the charge density. Therefore one can use on the same footing Hamiltonian theories
ranging from parameterized self-consistent tight-binding methods \cite{Alex}
to density functional theory \cite{DFT,KohnSham}. {\it Smeagol} uses DFT as its main
electronic structure method.

At this point it is important to observe that {\it Smeagol}, and
indeed any other NEGF DFT-based scheme, simply uses the Kohn-Sham
Hamiltonian \cite{KohnSham} as a single particle Hamiltonian. This
means that the non-equilibrium charge density obtained through the
NEGF method (equations (\ref{D_equil}) and (\ref{D_out_equil})) is
not by any mean associated with any variational principle and
certainly does not minimise the density functional, nor makes it
stationary. The only exception is for zero bias, where the method
presented here is just a clever alternative for solving an
equilibrium problem for an infinite non-periodic system. Although
it is common practice, it is therefore misleading and incorrect to
refer to our method as DFT-based NEGF, since the Hohenberg-Kohn
theorem cannot be applied \cite{DFT}. Some additional discussion
over this issue can be found in reference
[\onlinecite{smeagolsic}].

Although {\it Smeagol} is constructed in a simple and modular way and can be readily
interfaced with any DFT package based on a LAO basis set, for the present implementation
we have used the existing code {\it Siesta} \cite{siesta}.
{\it Siesta} is a mature numerical implementation of DFT, which has been specifically
designed for tackling problems involving a large number of atoms. It uses norm conserving
pseudopotentials in the separate Kleinman-Bylander form \cite{KB}, and most
importantly a very efficient LAO basis set \cite{MZ,SIESTAbasis1,SIESTAbasis2}.

\begin{figure}[ht]
\begin{center}
\includegraphics[width=7.0cm,clip=true]{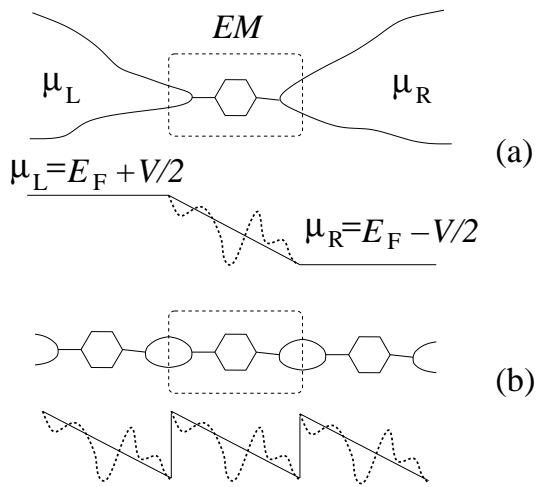}
\end{center}
\caption{\small{Schematic representation of the electrostatic problem. The real system
(a) of an extended molecule sandwiched between two leads is mapped onto a fictitious
periodic system (b), obtained by repeating the extended molecule in the direction of the
transport. The crucial point is that the potential profile in the unit cell of the periodic
system is identical to that of the actual structure.}}
\label{Fig3}
\end{figure}

One important aspect that deserves a mention is the way in which we calculate the Hartree
(electrostatic) potential for the extended molecule under bias. Clearly the easier and more
transparent way would be that of solving the Poisson's equation in real space with appropriate
boundary conditions. However this usually is numerically less efficient then solving it
in $k$-space using the fast Fourier transform algorithm. {\it Siesta} uses this second
strategy and so does {\it Smeagol}. In {\it Smeagol} the electrostatic potential is then
calculated for the infinite system obtained by repeating periodically the extended molecule
along the transport direction (see figure \ref{Fig3}). However, before solving the
Poisson's equation for such a system we add to the Hartree potential a saw-like term,
whose drop is identical to the bias applied. For convergence reasons we often add at
both edges of the scattering region two buffer layers, in which the external potential
is only a constant and the density matrix is that of the leads and is not
evaluated self-consistently.

In summary a typical {\it Smeagol} calculation proceeds as follows. First one computes
the leads self-energies over a range of energies $E$ as large as the bandwidth of the
materials forming the leads.
These are then stored either in memory or on disk depending on the size of the leads.
Then the proper {\it Smeagol} calculation is performed following the prescription described
in the previous sections.

\section{Test cases}\label{results}

We now present several test cases demonstrating the capabilities of {\it Smeagol}.
They address key aspects of the code such as the electrostatics, the calculation of the
transmission coefficients, the calculation of the $I$-$V$ characteristic, the
spin-polarization and the spin non-collinearity.

\subsection{Electrostatics: parallel plate capacitor}\label{capacitor}

As first simple test we present the case of a parallel plate capacitor, constructed
from two infinite fcc gold surfaces separated by a vacuum region 12.3~\AA\ long. Clearly
we do not expect transport across this device (with the exception of a tiny
tunneling current), but it is a good benchmark of the {\it Smeagol} ability
to describe the electrostatics of a device.

The two gold surfaces are oriented along the (100) direction and the unit cell has
only one atom in the cross section. The extended molecule comprises seven atomic planes
in the direction of the transport, which is enough for
achieving a good convergence of the Hartree potential (the Thomas-Fermi screening
length in gold is $\sim$0.6\AA\cite{Pettifor}). For the calculation we use 100
$k$-points in the full Brillouin zone in the transverse direction, a single
zeta basis set for the $s$, $p$ and $d$ orbitals and standard local density approximation
(LDA) of the exchange and correlation potential.

\begin{figure}[hbt]
\begin{center}
\includegraphics[width=8.0cm,clip=true]{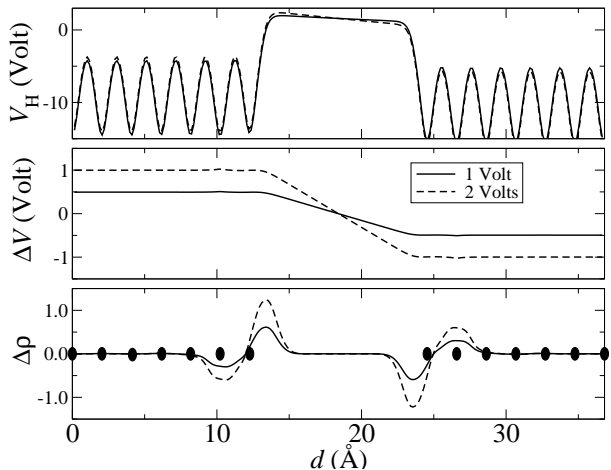}
\end{center}
\caption{a) Planar average of the Hartree potential $V_\mathrm{H}$ for an infinite
parallel-plate capacitor. b) Difference between the planar average of the Hartree
potential at a given bias and that at zero bias $\Delta V$. c) Difference $\Delta\rho$
between of the planar average of the charge density along the direction of the transport
for a given bias and that at zero bias. The dots indicates the position of gold planes.}
\label{Fig4}
\end{figure}

In figure \ref{Fig4} we present the planar average of the Hartree (electrostatic)
potential $V_\mathrm{H}$, the difference between the planar average of Hartree potential
at finite bias and that at zero bias $\Delta V$, and the difference $\Delta\rho$
between of the planar average of the charge density along the direction of the transport
for a given bias and that at zero bias.
The quantities shown in the picture are those expected from the
classical physics of a parallel plate capacitor. In the leads the
electrostatic potential shows oscillations with a period
corresponding to that of the separation between the gold planes,
but with a constant average. In contrast in the vacuum region the
potential is much higher, since there are no contributions from
the nuclei, but it is uniform. If we eliminate the oscillations,
by subtracting the zero bias potential from that obtained at
finite bias (figure \ref{Fig4}b), we obtain a constant potential
profile in the leads and a linear drop in the vacuum region.
Finally the macroscopic average of the charge density shows charge
accumulation on the surfaces of the capacitor and local charge
neutrality in the leads region as expected from a capacitor.

\subsection{Gold nanowires}

Metallic quantum point contacts (PCs) present conductance
quantization at room temperature \cite{ruitenbeek} a property that
has been predicted theoretically for many years\cite{GF4}.
Recently, Rodrigues {\it et al.} have shown that in a point
contact the crystallographic orientation of the atomic tips
forming the junction plays an important r\^ole in determining
transport properties \cite{Rodriguez}. Therefore, a realistic
theoretical description of the electronic transport in PCs must
take into consideration the atomistic aspects of the problem.

\begin{figure}[h]
\begin{center}
\includegraphics[width=8.5cm,clip=true]{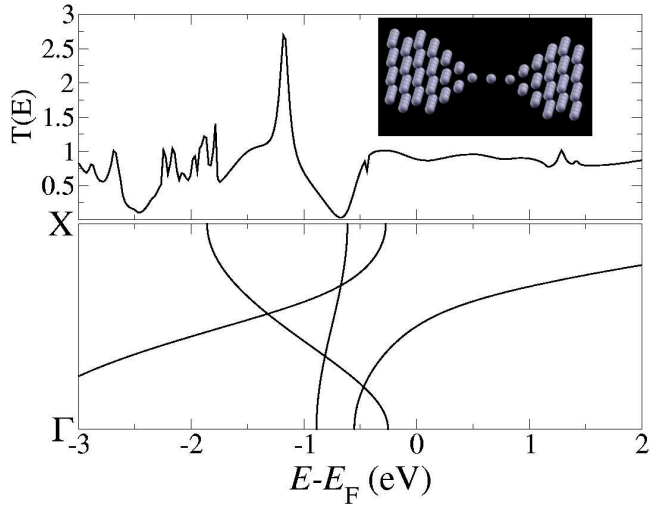}
\end{center}
\caption{The transmission coefficient as a function of energy
(upper panel) for a gold atomic point contact sandwiched between
two gold tips oriented along the [100] direction. In the lower
panel the bandstructure for a monoatomic gold chain with lattice
constant equal to the Au-Au separation in bulk gold. The inset
shows a ball-and-stick representation of the atomic positions of
the PC (the extended molecule).} \label{Fig5}
\end{figure}

As an example we have performed calculations for a [100]-oriented
gold quantum point contact (see inset of figure \ref{Fig5}). A
single gold atom is trapped at its equilibrium position between
two [100] fcc pyramids. This is the expected configuration for
such a specific crystal orientation, and the configuration likely
to form in breaking junction experiments for small elongation of
the junction. This has been confirmed by atomic resolution TEM
images \cite{taka,ugarte}. In this case we have used LDA and a
single zeta basis set for $s$, $p$ and $d$ orbitals. The unit cell
of the extended molecule now contains 141 atoms (seven planes of
the leads are included) and we consider periodic boundary
conditions with 16 $k$-points in the 2-D Brillouin zone.

In figure \ref{Fig5} we present the zero-bias transmission coefficient as a function of
energy. Recalling that the linear response conductance is simply $G=2e^2/h\:T(E_\mathrm{F})$
(in this case we have complete spin-degeneracy)
our calculation shows one quantum conductance for this point contact. Interestingly the
transmission coefficient is a rather smooth function $T\sim 1$ for a rather
broad energy range around $E_\mathrm{F}$. This means that the $G=2e^2/h$ result is
stable against the fluctuations of the position of the Fermi level, which may be
encountered experimentally.

The large plateau at $T\sim 1$ indicates the presence of a single
conductance channel for energies around and above $E_\mathrm{F}$.
This is expected from the bandstructure of a straight monoatomic
gold chain with lattice parameter equal to the Au-Au separation in
bulk gold (see figure \ref{Fig5}b), which presents only one $s$
band for such energy range. Therefore we conclude that the
transport at the Fermi level is dominated by a single
low-scattering $s$ channel. Notably for energies 1~eV below
$E_\mathrm{F}$ the transmission coefficient shows values exceeding
one, which are due to contributions from $d$ orbitals. In gold
mono-atomic chains these are substantially closer to
$E_\mathrm{F}$ than in bulk gold and participate to the transport.
These results are in good agreement with previously reported
calculations \cite{landman,rego_rocha} and experimental data
\cite{ruitenbeek,taka,RodriguezLett}. Additional examples of {\it
Smeagol}'s calculations for PCs carried out by the authors can be
found elsewhere in the literature \cite{vic1,vic2}.

\subsection{Molecular spin-valves}
The study of the $I$-$V$ characteristics of magnetic systems at the nanoscale is one
of {\it Smeagol}'s main goals. The most typical among spin devices is the magnetic
spin-valve, which is obtained by sandwiching a non-magnetic spacer between
two magnetic contacts. The direction of the magnetization in the two
contacts can be arbitrarily changed by applying a magnetic field. The device
then switches from a low resistance state, when the magnetization vectors in the leads
are parallel to each other, to a high resistance state when the alignment of the
magnetizations is antiparallel. This is the giant magnetoresistance effect \cite{GMR1,GMR2},
which is at the foundation of modern hard-disk reading technology.

Traditional spin-valves use either metals or inorganic insulators
as spacers. However a recent series of experiments have shown that
organic molecules can serve the same purpose and a rather large
GMR can be found \cite{bruce,Aws1,Shi,Dediu,Ralph}. These
experiments could lead to integrating the functionalities of
molecules with spin-systems, and therefore have the potential to
merge together the fields of spin- and molecular-electronics.

The calculation of the transport properties of molecular
spin-valves is a tough theoretical problem. It involves the
computation of accurate electronic structures for magnetic
surfaces, the charging properties of molecules, and the knowledge
of the actual atomic positions. In a recent paper \cite{Smeagol2}
we have demonstrated that molecules can efficiently be employed in
spin-valves. Moreover we have shown that $\pi$-conjugated
conducting molecules produce larger GMR then their insulating
counterparts. Most of the effect is due to the orbital selectivity
of the molecule/metal bonding, which in transition metals
translates to a spin-selectivity. Here we further expand this
concept and we demonstrate that the GMR can be tuned by molecular
end-group engineering.
\begin{figure}[h]
\begin{center}
\includegraphics[width=7.5cm,clip=true]{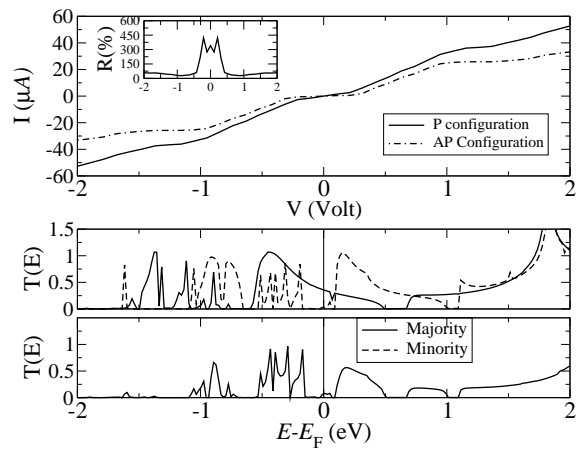}
\end{center}
\caption{Transport properties for a 1,4-phenyl molecule attached to Ni (100) surfaces
through a S group. The top panel shows the $I$-$V$ characteristics for both the parallel and
antiparallel alignment of the leads and the inset the corresponding GMR ratio. The lower
panel is the transmission coefficient at zero bias as a function of energy. Because of
spin-symmetry, in the antiparallel case we plot only the majority spin.}
\label{Fig6}
\end{figure}

The system under investigation is a 1,4-phenyl molecule attached
to two fcc Ni surfaces oriented along the (001) direction. The
molecules are attached to the Ni hollow site through a thiol-like
group where we use S, Se and Te as anchoring atoms. We consider
collinear spin only and investigate the $I$-$V$ characteristic
assuming the magnetization vectors in the current/voltage contacts
to be either parallel (P) or antiparallel (AP) to each other. The
size of the GMR effect is expressed by the GMR ratio
$R_\mathrm{MR}$, which is defined as
$R_\mathrm{MR}=(I_\mathrm{P}-I_\mathrm{AP})/I_\mathrm{AP}$, with
$I_\mathrm{P}$ ($I_\mathrm{AP}$) the current in the parallel
(antiparallel) state. At zero bias, when all the currents vanish,
we replace them with the conductances.

We construct the unit cell of the extended molecule to include
four Ni atomic planes on each side, for a total of forty Ni atoms.
The basis set is critical and a single zeta for all the orbitals
is not sufficient. Therefore we have used single zeta for H, C and
S $s$ orbitals, double zeta for Ni $s$, $p$ and $d$, and double
zeta polarized for C and S $p$ orbitals. This basis gives us a
Hamiltonian with over a thousand degrees of freedom. Finally the
charge density is obtained by integrating the Green's function
over 50 imaginary and 600 real energies.

\begin{figure}[h]
\begin{center}
\includegraphics[width=7.5cm,clip=true]{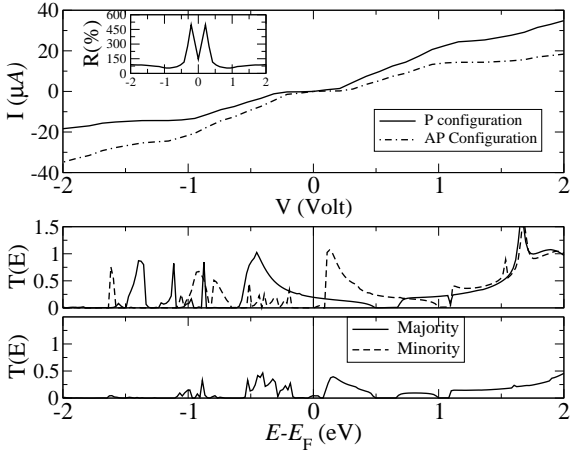}
\end{center}
\caption{Transport properties for a 1,4-phenyl molecule attached to Ni (100) surfaces
through a Se group. The top panel shows the $I$-$V$ characteristics for both the parallel and
antiparallel alignment of the leads and the inset the corresponding GMR ratio. The lower
panel is the transmission coefficient at zero bias as a function of energy. Because of
spin-symmetry, in the antiparallel case we plot only the majority spin.}
\label{Fig7}
\end{figure}

\begin{figure}[h]
\begin{center}
\includegraphics[width=7.5cm,clip=true]{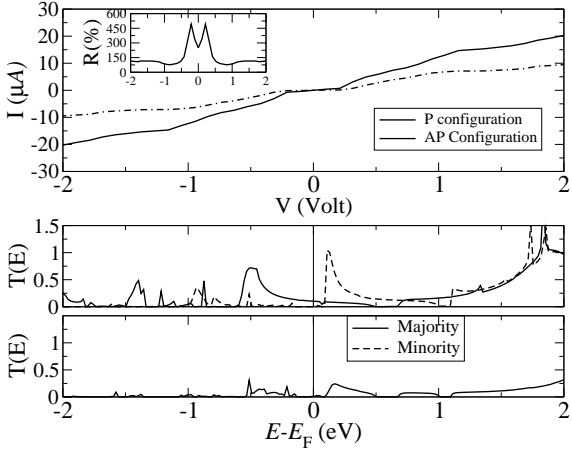}
\end{center}
\caption{Transport properties for a 1,4-phenyl molecule attached to Ni (100) surfaces
through a Te group. The top panel shows the $I$-$V$ characteristics for both the parallel and
antiparallel alignment of the leads and the inset the corresponding GMR ratio. The lower
panel is the transmission coefficient at zero bias as a function of energy. Because of
spin-symmetry, in the antiparallel case we plot only the majority spin.}
\label{Fig8}
\end{figure}

In figures \ref{Fig6}, \ref{Fig7}  and \ref{Fig8} we present the
$I$-$V$ characteristics, the zero bias transmission coefficient as
a function of energy, and the GMR ratio as a function of bias for
the three anchoring situations (S, Se, Te). Clearly all the three
cases show a large GMR, particularly for small biases.
Interestingly the maximum GMR increases when going from S to Se to
Te, and this is correlated with a general reduction of the total
transmission and consequently of the current. Such a reduction is
more pronounced in the case of antiparallel alignment of the leads
and this gives rise to the increase in GMR. The origin of the
drastic reduction of the transmission when changing the anchoring
groups has to be found in the different bonding structure. Since
S, Se and Te all belong to the same row of the periodic table, the
orbital nature of the bonding to the Ni surface is left unchanged,
and so are the generic features of the transmission coefficient.
However the bond distance goes from 1.28\AA\ to 1.48\AA\ to
1.77\AA\ when going from S to Se to Te. This large increase in the
bond distance is responsible for the reduction in transmission.

The most relevant features of the transmission coefficient can be understood in terms of
tunneling through a single molecular state \cite{Smeagol2}. If we define $t^\uparrow(E)$
($t^\downarrow(E)$) as the majority (minority) spin hopping integral from one of the
leads to the molecular state, then the total transmission coefficient through the entire
device in the parallel alignment will be simply $T=T^{\uparrow\uparrow}+T^{\downarrow\downarrow}=
(t^{\uparrow})^2+(t^{\downarrow})^2$. Here the total transmission coefficient for
majority (minority) spin is $T^{\uparrow\uparrow}=(t^{\uparrow})^2$
($T^{\downarrow\downarrow}=(t^{\downarrow})^2$). Similarly in the
case of antiparallel alignment of the leads we have $T=2T^{{\uparrow\downarrow}}=
2T^{\downarrow\uparrow}=2t^{\uparrow}t^{\downarrow}$.
In this simple description, which neglects both co-tunnelling and multiple scattering from the
contacts, $T^{{\uparrow\downarrow}}(E)$ turns out to be a convolution of the transmission
coefficients for the parallel case $T^{\uparrow\downarrow}\propto
\sqrt{{T^{\uparrow\uparrow}}T^{\downarrow\downarrow}}$. This type of behavior can
be appreciated in figures \ref{Fig6}, \ref{Fig7} and \ref{Fig8}.

It is important to note that the large GMR ratio is ultimately due
to the low transmission around $E_\mathrm{F}$ in the antiparallel
case, which originates from the small transmission of the minority
spins in the parallel case through the relation
$T^{\uparrow\downarrow}(E_\mathrm{F})\propto
\sqrt{{T^{\uparrow\uparrow}}(E_\mathrm{F})T^{\downarrow\downarrow}(E_\mathrm{F})}$.
This is surprising since the density of states for minority spins
at the Fermi level is rather large and one may expect substantial
transmission. Moreover $s$-like electrons, which are weakly
affected by the spin orientation, generally contribute heavily to
the current regardless of the magnetic state of the device.
Therefore, what does block the minority spin electrons?

A detailed analysis of the local density of states of the molecule attached to the leads
\cite{Smeagol2} reveals that the bonding of the molecule to the Ni surface is
through Ni $d$ and S $p$ orbitals (the same is valid for Se and Te). The transport is
therefore through hybrid Ni~$d$-S~$p$ states, which in turns are spin-split due to the
ferromagnetism of Ni. The crucial point here is that for the minority band these states
end up above the Fermi level, and therefore do not contribute to the low bias
transport. This is an important observation, since it demonstrates that orbital
selectivity in magnetic systems can produce a spin-selectivity, and therefore
magnetoresistance type of effects.

\subsection{Nickel point contacts}

The transport properties of magnetic transition metal point
contacts have been the subject of several recent investigations.
Technologically these systems are attractive since they can be
used as building blocks for read heads in ultra-high density
magnetic data storage devices. From a more fundamental point of
view they offer the chance to investigate magnetotransport at the
atomic level. Magnetic point contacts are effectively
spin-valve-like devices, with the spacer now replaced by a narrow
constriction where a sharp domain wall can nucleate \cite{bruno}.
Therefore the magnetoresistance can be associated with domain-wall
scattering and the MR ratio can be defined earlier.

A simple argument based on the assumption that all the valence
electrons can be transmitted with $T\sim 1$ gives an upper bound
for the GMR of the order of a few percent (100\% in the case of
nickel). This however may be rather optimistic since one expects
the $d$ electrons to undergo quite some severe scattering. Indeed
small values of GMR for Ni point contacts have been measured
\cite{viret1}. Surprisingly at the same time other groups have
measured huge GMR for the same system
\cite{PC1,garcia2001,zchopra,oscar3}. Although mechanical effects
can be behind these large values \cite{viret2}, the question on
whether or not a large GMR of electronic origin can be found in
point contacts remains.

\begin{figure}[h]
\begin{center}
\includegraphics[width=8.5cm,clip=true]{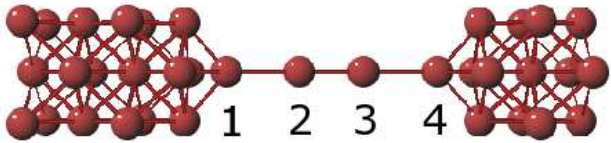}
\end{center}
\caption{Schematic representation of the Ni point contact simulated. In the symmetric case
the domain wall is located between the second and the third atom, while in the asymmetric
it is placed between the third and the fourth. The direction of the current is from 1 to 4
for positive bias.}
\label{Fig9}
\end{figure}

Therefore we investigate the zero bias conductance of a four atom
long monoatomic Ni chain sandwiched between two Ni (001) surfaces
(see figure \ref{Fig9}). This is an extreme situation rarely found
in actual break junctions \cite{Bahn}. However an abrupt domain
wall (one atomic spacing long) in a monoatomic chain is the
smallest domain wall possible, and it is expected to show the
larger GMR. For this reason our calculations represent an upper
bound on the GMR obtainable in Ni only devices, and they also
serve also as a test of the {\it Smeagol} capability for dealing
with non-collinear spin.

In this calculation we use a double zeta basis set for $s$, $p$
and $d$ orbitals and consider finite leads (no periodic boundary
conditions are applied) with either four or five atoms in the
cross section. We then investigate two possible situations. In the
first one we place the domain wall symmetrically with respect to
the leads, i.e. between the second and the third atom of the
chain. In the second (asymmetric) the domain wall is positioned
between the third and the fourth atom. Furthermore we perform
spin-collinear and spin-non-collinear calculations for both cases.
Interestingly all our non-collinear calculations always converge
to a final collinear solution. This confirms expectations based on
simple $s$-$d$ model \cite{Maria}, suggesting that the strong
exchange coupling between the conduction electrons and those
responsible for the ferromagnetism, stabilize the collinear state
if the magnetization vectors of the leads are collinear.

\begin{figure}[h]
\begin{center}
\includegraphics[width=8.5cm,clip=true]{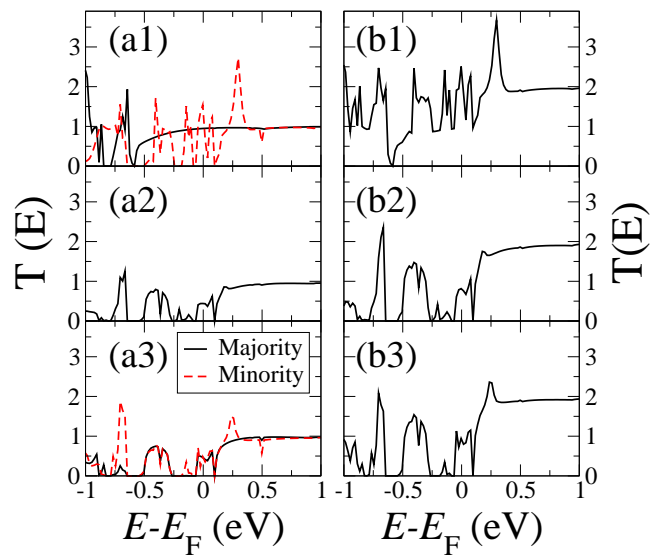}
\end{center}
\caption{Transmission coefficient as a function of energy for the nickel quantum point contacts
of figure \ref{Fig9}. The right-hand side panels (a) are for collinear calculations and
the left-hand-side (b) are for non-collinear; (1) parallel state, (2) antiparallel with
symmetric domain wall, (3) antiparallel with asymmetric domain wall. Note that in the
non-collinear case we do not distinguish between majority and minority spins. In panel (a2)
majority and minority spins are degenerate.}
\label{Fig10}
\end{figure}

In figure \ref{Fig10} we present the transmission coefficient as a function of the
energy for both the symmetric and asymmetric case and the parallel
state. For collinear calculations the contributions from majority and minority spins are
plotted separately, while we have only one transmission coefficient in
the non-collinear case. Clearly in all cases the non-collinear
solution agrees closely with the collinear one, i.e. $T_\mathrm{collinear}^\uparrow+
T_\mathrm{collinear}^\downarrow=T_\mathrm{non-collinear}$. This is expected since the final
magnetic arrangement of the non-collinear calculation is actually collinear, and it
is a good test for our computational scheme.

Turning our attention to the features of the transmission coefficient
it is evident that at the Fermi level $T$ in the parallel state is larger
than that in the antiparallel. This difference however is not large and
the GMR ratio is about 60\% with little
difference between the symmetric and asymmetric domain wall. This is mainly due to
the much higher transmission of the un-polarized $s$ electrons compared with
that of the $d$. Note that the conductance approaches $2e^2/h$ for energies approximately
0.5~eV above the Fermi level. For such energies in fact no $d$ electrons contribute to
the density of states of both the spin sub-bands, and only $s$ electrons are left. These are
then transmitted with $T\sim 1$ as in the case of Au chains investigated previously.

The crucial point is that the contribution of the $s$ electrons is
also large at the Fermi level. This results in a poorly
spin-polarized current at low bias and consequently in a small
GMR, in agreement with other calculations \cite{mertig,palacios}.
In conclusion our finding rules against the possibility of large
GMR from electronic origin in Ni point contacts. However the
presence of non-magnetic contamination (for example oxygen) may
change this picture radically.

\subsection{H$_2$ molecules joining platinum electrodes}

The aim of this section is to show how strongly the transmission
coefficients may depend on the leads cross section, whenever
$d$ electrons are close to $E_\mathrm{F}$.
As an example, we present results for the H$_2$ molecule sandwiched
between fcc Pt(001) leads, and compare leads of different sections with extended
leads. These are obtained by applying periodic boundary conditions and a sampling
over the $k$-points along the direction orthogonal to that of transport.

The conductance of an H$_2$ molecule sandwiched between platinum
electrodes has been extensively studied
\cite{smit,cuevas,garcia-pala,thygesen,vic1}. Experimentally
\cite{smit} it has been found that the inclusion of hydrogen gas
into the vacuum chamber produces a dramatic change in the
conductance histograms of platinum, which change from a structure
with a broad peak at 1.5 $G_0$ to a structure with a sharp peak at
1 $G_0$. This resonance has been attributed to the conductance
through a single molecule, which bridges both leads lying parallel
to the current flow. This explanation has been confirmed by
theoretical calculations \cite{cuevas,thygesen,vic1}.
\begin{figure}[h]
\begin{center}
\includegraphics[width=8.5cm,clip=true]{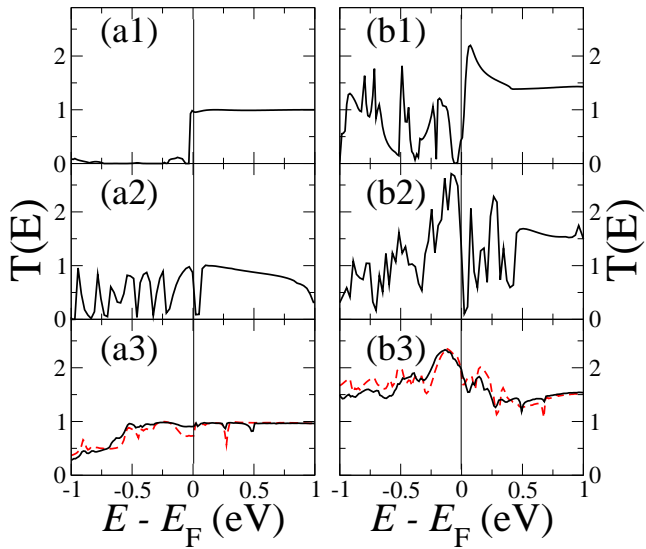}
\end{center}
\caption{Transmission coefficients for an H$_2$ molecule
sandwiched between fcc platinum leads near the equilibrium
distances ($\sim$ 9.5 and 11 \AA). The left hand side (a)
corresponds to the configuration where the molecule lies parallel
to the current flow and the right hand side (b) to the
configuration where the molecule lies perpendicular. The leads are
made of alternating slabs of 4-5 atoms (1) and 9-12 atoms (2)
without periodic boundary conditions along the perpendicular
directions ($xy$), and 9-9 atoms with periodic boundary conditions
along $xy$ (3). In the last case the dashed and
continuous lines have been obtained with 4 and 12 $k$ points,
respectively.} \label{Fig11}
\end{figure}

In our calculations the H$_2$ molecules is located either parallel
or perpendicular to the current flow. A detailed description of
the geometric configuration and the results can also be found in
reference [\onlinecite{vic1}]. We use a double zeta polarized
basis set for platinum $s$, $p$ and $d$ orbital, a double zeta for
the hydrogen $s$ electrons and the LDA functional. As a first step
we employ finite cross section leads along the transversal
directions, composed of alternated planes containing 4 and 5 atoms
each. The resulting transmission coefficients show many peaks and
gaps throughout all the energy range and particularly sharp
variations around the Fermi energy, as can be seen in figures
\ref{Fig11} (a1) and (b1). When thicker slabs composed of
alternating planes of 9 and 12 atoms are employed the results do
not improve and the large oscillations still remain, as shown in
figures \ref{Fig11} (a2) and (b2). It is apparent from these
figures that while $T(E)$ shows a long plateau at positive
energies, it presents strong oscillations at the Fermi energy and,
therefore, it is uncertain to infer the conductance of the
junction from $T(E_\mathrm{F}$).

This is in stark contrast with the case of gold, where the $d$-levels lie below
$E_\mathrm{F}$, and $T(E)$ is smooth regardless of the size of the leads cross
section. For platinum the presence of $d$-states at the Fermi energy opens
minibands and minigaps, which translate into strong oscillations in
$T(E\sim E_\mathrm{F})$. These minibands and minigaps arise from
interference effects of the $d$-states along the transverse
direction. Consequently, oscillations in $T(E)$ should disappear
when bulk electrodes are used. Indeed, this is what we find when
slabs made of 3$\times$3 atomic planes and periodic boundary conditions
are employed, as shown in figures \ref{Fig11} (a3) and (b3). We moreover show how
$T(E)$ converges when the number of transverse $k$ points is
increased from 4 to 12. Although some small variations and peaks
still remain when 4 $k$ points are used, the transmission at the
Fermi level is essentially converged. Note that the parallel case
has $T\sim 1$ for a long range of energies around $E_{\mathrm F}$,
which remains essentially unperturbed for small variations of the
coordinates or the distance between the electrodes. This explains
the sharp peak observed in the experimental conductance histograms
\cite{smit}.

In view of the above calculations we can therefore conclude that
the use of bulk electrodes, characterized by periodic boundary
conditions along the perpendicular directions and $k$ points, is
mandatory in order to avoid oscillations in the transmission
coefficients. Otherwise the presence of strong variations and
minigaps can give unphysical solutions for systems with open $d$
shells.

\section{Conclusions}

We have presented a description of our newly developed
non-equilibrium Green's function code {\it Smeagol}. In the
present version {\it Smeagol} uses the DFT implementation
contained in {\it Siesta} as the underlying electronic structure
method. However the code has been developed in a modular and
general form and can be easily combined with any electronic
structure scheme based on localized orbital basis set. The core of
{\it Smeagol} is our new algorithm for calculating the surface
Green's functions of the leads, which combines generalized
singular value decomposition with decimation. This results in an
un-precedent numerical stability for a quantum transport code, and
in the possibility of drastically reducing the number of degrees
of freedom in the leads. In this way large current/voltage probes
with complicated electronic structure can be tackled.

We have also presented a selection of results obtained with {\it Smeagol}. These range
from simple tests for the electrostatics to an analysis of the GMR in molecular
spin-valves, and demonstrate the {\it Smeagol}'s ability to tackle very
different problems.

\section*{Acknowledgement}

The authors thank K.~Burke and T.~Todorov for discussion.
This work is sponsored by Science Foundation of Ireland under the
grant SFI02/IN1/I175, the UK EPSRC and the EU network
MRTN-CT-2003-504574 RTNNANO. JF and VMGS thank the Spanish
Ministerio de Educaci\'on y Ciencia for financial support (grants
BFM2003-03156 and AP2000-4454). ARR thanks Enterprise Ireland
(grant EI-SC/2002/10) for financial support. Traveling has been
sponsored by the Royal Irish Academy under the International
Exchanges Grant scheme.

\appendix
\section{Stable algorithm for the evaluation of the self-energies}\label{H1problem}
\subsection{The ``$K_1$ problem''}
The method presented in section \ref{leads} to calculate the leads Green's functions depends
crucially on the fact that the coupling matrix between principal layers $K_1=H_1-ES_1$ is
invertible and not ill-defined. However this is not necessarily the case since
singularities can be present in $K_1$ as the result of poor coupling between
PLs or because of symmetry reasons. Note also that since
$K_1=H_1-ES_1$ the rank of $K_1$ may also depend on the energy $E$.

\begin{figure}[h]
\begin{center}
\includegraphics[width=6.5cm,clip=true]{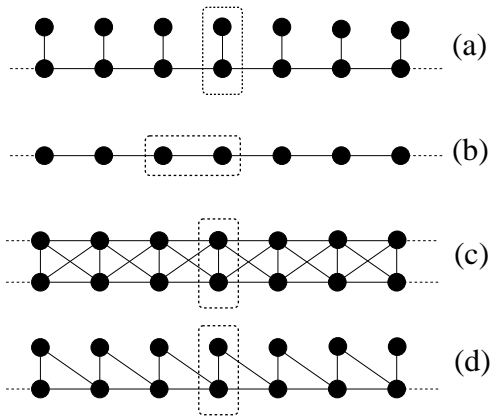}
\end{center}
\caption{Four different structures for which $H_1$ is singular: (a) lack of bonding,
(b) supercell, (c) over-bonding, (d) odd bonding. Each black dot represents an
atom and each line a bond. The dashed boxes enclose a principal layer.}
\label{FigA1}
\end{figure}

We now give a few examples illustrating how these singularities arise. Let us consider
for the sake of simplicity an orthogonal nearest neighbour tight-binding model with
only one $s$-like basis function per atom. In this case $K_1=H_1$ is independent from
the energy. In figure \ref{FigA1} we present four possible cases for which
$H_1$ is singular.
In the picture the dots represent the atomic position, the lines the bonds and the
dashed boxes enclose a PL. All the bonds are assumed to have the same
strength, thus all hopping integrals $\gamma$ are identical.

In the first case (figure \ref{FigA1}a) the PL coincides with the primitive
unit cell of the system and therefore it is the smaller principal layer that can be
constructed. However since every second atom in the cell does not couple with its
mirror in the two adjacent cells $H_1$ has the form
\begin{equation}
H_1=
\left(\begin{array}{cc}
\gamma & 0 \\
0 & 0 \\
\end{array}\right)\:
\label{h11}
\end{equation}
and therefore is singular. This is the case of ``lack of bonding'' between principal layers.
It is the most common case and almost always present when dealing with transition metals,
since localized $d$ shells coexist with delocalized $s$ orbitals.

Figure \ref{FigA1}b presents a different possibility. Here the PL is a
supercell constructed from two unit cells and every atom in the PL couples
with atoms located in only one of the two adjacent PLs. In this case
\begin{equation}
H_1=
\left(\begin{array}{cc}
0 & 0 \\
\gamma & 0 \\
\end{array}\right)\:,
\label{h12}
\end{equation}
which is again singular. Clearly in this specific case one can reduce the principal
layer to be the primitive unit cell solving the problem ($H_1$ become a scalar $\gamma$).
However in a multi-orbital scheme the supercell drawn may be the smallest PL
possible and the problem will appear. Again this is a rather typical situation
when dealing with transition metals.

The case of ``overbonding'' is shown in figure \ref{FigA1}c. Again the
PL coincides with the primitive unit cell, but now every atom in the
PL is coupled to all the atoms in the two adjacent PLs.
In this case
\begin{equation}
H_1=
\left(\begin{array}{cc}
\gamma & \gamma \\
\gamma & \gamma \\
\end{array}\right)\:,
\label{h13}
\end{equation}
which is not invertible.

Finally the ``odd bonding'' case is presented in figure \ref{FigA1}d. Also in this case the PL
coincides with the primitive unit cell, however the upper atom in the cell is coupled
only to atoms in the right nearest neighbour principal layer. The $H_1$ matrix is then (we label
as ``1'' the upper atom in the cell)
\begin{equation}
H_1=
\left(\begin{array}{cc}
 0 & \gamma \\
 0 & \gamma \\
\end{array}\right)\:,
\label{h14}
\end{equation}
i.e. it is singular. Clearly the above categorization is basis
dependent, since one can always find a unitary rotation
transforming a generic $H_1$ in a new matrix of the form of
equation (\ref{h14}).

\subsection{Finding the singularities of $K_1$}\label{appendixA1}

We now present the first step of a scheme for regularizing $K_1$,
and indeed the whole Hamiltonian and overlap matrix, by removing
their singularities. In the cases of ``lack of bonding'',
``supercell'' and ``odd bonding'' presented in the previous
section the singularities of $K_1=H_1$ were well defined since an
entire column was zero. However more generally, and in particular
in the case of multiple zetas basis set, $K_1$ is singular without
having such a simple structure (for instance as in the
``overbonding'' case). This is the most typical situation and a
method for identifying the singularities is needed.

The ultimate goal is to perform a unitary transformation of both ${\cal H}$ and
${\cal S}$ in such a way that the off-diagonal blocks of the leads Hamiltonian
and overlap matrix ($H_1$ and $S_1$) assume the form
\begin{equation}
\begin{array}{cc}
&  \scriptstyle{N-R} \ \ \scriptstyle{R} \\
\scriptstyle{N} &
[\:0\:\:,\:\: A\: ]
\end{array}
=
\left(
\begin{array}{cccccc}
0 & 0 & \cdots & A_{1,N-R+1} & \cdots & A_{1,N} \\
0 & 0 & \cdots & A_{2,N-R+1} & \cdots & A_{2,N} \\
0 & 0 & \cdots & A_{3,N-R+1} & \cdots & A_{3,N} \\
\vdots & \vdots & \vdots & \vdots & \vdots & \vdots \\
0 & 0 & \cdots & A_{N,N-R+1} & \cdots & A_{N,N} \\
\end{array}
\label{blockform}
\right)\:,
\end{equation}
i.e. they are $N\times N$ block matrices of rank $R$, whose first $N-R$
columns vanish. In this form the problem is re-conducted to the problem of
``odd bonding'' presented in the previous section.

This can be achieved by performing a generalized singular value decomposition (GSVD)
\cite{lapack}. The idea is that a pair of $N\times N$ matrices, in this case $H_1$
and $S_1$, can be written in the following form
\begin{eqnarray}
H_{1}= U\Lambda_1\left[0,W\right]Q^\dagger, \\
S_{1}= V\Lambda_2\left[0,W\right]Q^\dagger,
\end{eqnarray}
with $U$, $V$ and $Q$ unitary $N\times N$ matrices and $W$ being
a $R \times R$ non-singular triangular matrix where $R$ is the rank of the
$2N\times N$ matrix $\left[\begin{array}{c} H_{1}\\ S_{1} \end{array}\right]$
($R \leq N$). The matrices $\Lambda_1$ and $\Lambda_2$ are defined as follows:
\begin{equation}
\Lambda_1=
\begin{array}{cc}
 & \scriptstyle{K} \ \ \scriptstyle{L} \\
\begin{array}{r}
\scriptstyle{K} \\
\scriptstyle{L} \\
\scriptstyle{N - K - L}
\end{array} &
\left(\begin{array}{cc}
I_K & 0 \\
0 & C \\
0 & 0 \end{array}\right)
\end{array},
\end{equation}
\begin{equation}
\Lambda_2=
\begin{array}{cc}
 & \scriptstyle{K} \ \ \scriptstyle{L} \\
\begin{array}{r}
\scriptstyle{L} \\
\scriptstyle{N - L}
\end{array} &
\left(\begin{array}{cc}
0 & C^\prime \\
0 & 0 \end{array}\right)
\end{array}
\end{equation}
where $L$ is the rank of $S_1$, $K+L=R$, $I_K$ is the $K\times K$ unit matrix
and $C$ and $C^\prime$ are matrices to determine.

Clearly the two matrices $H_1$ and $S_1$ have the two common generators
$W$ and $Q$. Then, one can perform a unitary transformation of
both $H_1$ and $S_1$ by using $Q$, obtaining
\begin{eqnarray}
H_1^Q=Q^\dagger H_1Q = Q^\dagger U\Lambda_1\left[0,W\right]=
\begin{array}{rc}
  & \scriptstyle{N-R} \ \ \ \scriptstyle{R} \\
\scriptstyle{N} &
\left[
\begin{array}{ccc}
0 &  , & \bar{H}_{1}
\end{array}\right]
\end{array}\label{h1bar}
\\
S_1^Q=Q^\dagger S_1Q = Q^\dagger V\Lambda_2\left[0,W\right]=
\begin{array}{cc}
  & \scriptstyle{N-R} \ \ \ \scriptstyle{R} \\
\scriptstyle{N} &
\left[\begin{array}{ccc}
0 & , & \bar{S}_{1}
\end{array}\right].
\end{array}
\end{eqnarray}
Here $\bar{H}_{1}$ and $\bar{S}_{1}$ are the $N\times R$ non-vanishing blocks of the
GSVD transformed matrices $H_1$ and $S_1$ respectively.

In an analogous way the same transformation for $H_1^\dagger$, $S_1^\dagger$, $H_0$ and $S_0$
leads to
\begin{eqnarray}
{H^Q_1}^\dagger=Q^\dagger H_1^\dagger Q =
\begin{array}{cc}
  & \scriptstyle{N} \\
\begin{array}{c}
\scriptstyle{N-R} \\
\scriptstyle{R}
\end{array} &
\left[\begin{array}{c}
0 \\
\bar{H}_{1}^\dagger
\end{array}\right]
\end{array}
\\
{S^Q_1}^\dagger=Q^\dagger S_1^\dagger Q =
\begin{array}{cc}
  & \scriptstyle{N} \\
\begin{array}{c}
\scriptstyle{N-R} \\
\scriptstyle{R}
\end{array} &
\left[\begin{array}{c}
0 \\
\bar{S}_{1}^\dagger
\end{array}\right]
\end{array}
\\
H_0^Q=Q^\dagger H_0Q\:,
\\
S_0^Q=Q^\dagger S_0Q\:,
\end{eqnarray}
where the transformed matrices $H_0^Q$ and $S_0^Q$ are not necessarily
in the form of equation (\ref{blockform}).

We are now in the position of writing the final unitary transformations for the total
(infinite) Hamiltonian ${\cal H}$ and overlap ${\cal S}$ matrices describing
the whole system (leads plus extended molecule). These are given by
${\cal Q}^\dagger{\cal HQ}$ and ${\cal Q}^\dagger{\cal SQ}$ with the
infinite matrix ${\cal Q}$ defined as
\begin{equation}
{\cal Q}
=
\left(\begin{array}{ccccccccc}
\ddots&.&.&.&.&.&.&.&\\
.&0&Q&0&.&.&.&.&.\\
.&.&0&Q&0&.&.&.&.\\
.&.&.& 0& I_\mathrm{M} &0&.&.&.\\
.&.&.&.&0&Q&0&.&.\\
.&.&.&.&.&0&Q&0&.\\
.&.&.&.&.&.&.&.&\ddots\\
\end{array}\right)\;,
\end{equation}\label{Qtransf}
where $I_\mathrm{M}$ is the $M\times M$ unit matrix. Note that this unitary
transformation rotates all the $H_1$ matrices (the $S_1$ matrices in the case of
${\cal S}$), but it leaves $H_\mathrm{M}$ ($S_\mathrm{M}$) unchanged. Finally the
matrices coupling the extended molecule to the leads transform as following
\begin{equation}
\begin{array}{l}
H_\mathrm{LM}^Q\:\rightarrow\:Q^\dagger H_\mathrm{LM}\:, \\
H_\mathrm{ML}^Q\:\rightarrow\: H_\mathrm{ML}Q\:, \\
H_\mathrm{RM}^Q\:\rightarrow\:Q^\dagger H_\mathrm{RM}\:, \\
H_\mathrm{MR}^Q\:\rightarrow\: H_\mathrm{MR}Q\;, \\
\end{array}
\end{equation}
and so do the corresponding matrices of ${\cal S}$.

\subsection{Solution of the ``$K^1$ problem''}\label{appendixA2}

Now that both ${\cal H}$ and ${\cal S}$ have been written in a convenient form
we can efficiently renormalize them out. The key observation is that the two
(infinite) blocks describing the leads have now the following structure
(the ${\cal S}$ matrix has an analogous structure and it is not shown here
explicitly)
\begin{widetext}
\begin{equation}\label{h1problem}
{\cal H}_\mathrm{L/R}^Q=
\left(\begin{array}{ccccc}
\ddots & \vdots & \vdots & \vdots &  \\
\cdots & Q^\dagger{H}_0Q    & Q^\dagger{H}_1Q    & 0      & \cdots \\
\cdots & Q^\dagger{H}_{-1}Q & Q^\dagger{H}_0Q    & Q^\dagger{H}_1Q    & \cdots \\
\cdots & 0      & Q^\dagger{H}_{-1}Q   & Q^\dagger{H}_0Q    & \cdots \\
       & \vdots & \vdots & \vdots & \ddots
\end{array}\right)
=
\left(\begin{array}{ccccc}
\ddots & \vdots & \vdots & \vdots &  \\
\cdots & Q^\dagger{H}_0Q    & \left[\begin{array}{cc}
0 & \bar{H}_1 \\
\end{array}\right]      & 0      & \cdots \\
\cdots & \left[\begin{array}{c}
0 \\
\bar{H}_1^\dagger
\end{array}\right]  &
\left(\begin{array}{cc}
C & B \\
B^\dagger & D
\end{array}\right)  &
\left[\begin{array}{cc}
0 & \bar{H}_1 \\
\end{array}\right]    & \cdots \\
\cdots & 0      &
\left[\begin{array}{c}
0 \\
\bar{H}_1^\dagger
\end{array}\right] &
\left(\begin{array}{cc}
C & B \\
B^\dagger & D
\end{array}\right)    & \cdots \\
       & \vdots & \vdots & \vdots & \ddots
\end{array}\right)\:,
\label{matform}
\end{equation}
\end{widetext}
where the matrices $D$, $B$ and $C$ are respectively $R\times R$,
$N\times(N-R)$ and $(N-R)\times(N-R)$.

Note that the degrees of freedom (orbitals) contained in the block $C$ of the matrix
$H_0^Q=Q^\dagger{H}_0Q$ couple to those of only one of the two adjacent PLs.
This situation is the generalization to a multi-orbital non-orthogonal tight-binding model
of the ``odd bonding'' case discussed at the beginning of this appendix (figure \ref{FigA1}d).
These degrees of freedom are somehow redundant and they
will be eliminated. We therefore proceed with performing Gaussian elimination \cite{rgf}
(also known as ``decimation'') of all the degrees of freedom associated to all the blocks
$C$.

The idea is that the Schr\"odinger equation ${\cal Q}^\dagger[{\cal H}-E{\cal S}]{\cal Q}\Psi=0$ can be
re-arranged in such a way that a subset of degrees of freedom (in this case those associated
to orbitals in a PL that couple only to one adjacent PL) do not appear explicitly.
The procedure is recursive. Let us suppose we wish to eliminate the $l$-th row and
column of the matrix ${\cal K}^Q={\cal Q}^\dagger[{\cal H}-E{\cal S}]{\cal Q}$. This can be done by
re-arranging the remaining matrix elements according to
\begin{equation}
{{\cal K}^Q}^{(1)}_{ij}={{\cal K}^Q}_{ij} -
\frac{{{\cal K}^Q}_{il}{{\cal K}^Q}_{lj}}{{{\cal K}^Q}_{ll}}\:.
\end{equation}
The dimension of the resulting new matrix ${{\cal K}^Q}^{(1)}$ (``1'' indicates that
one decimation has been performed) is reduced by one with respect to the original
${\cal K}^Q$. This procedure is then repeated and after $r$ decimations we obtain a
matrix
\begin{equation}
{{\cal K}^Q}^{(r)}_{ij}={{\cal K}^Q}^{(r)}_{ij} -
\frac{{{\cal K}^Q}^{(r-1)}_{il}{{\cal K}^Q}^{(r-1)}_{lj}}
{{{\cal K}^Q}^{(r-1)}_{ll}}\;.
\end{equation}

Let us now decimate all the matrix elements contained in all the sub-matrices $C$.
We obtain a new tridiagonal matrix ${{\cal K}^Q}^{(\infty)}$ (``$\infty$'' means that
an infinite number of decimations have been performed) of the form
\begin{equation}
{{\cal K}^Q}^{(\infty)}
=
\left(\begin{array}{ccccccccccc}
.&.&.&.&.&.&.&.&.&.&\\
0&\Theta^\dagger&\Delta&\Theta&0&.&.&.&.&.&.\\
.&0&\Theta^\dagger&\Delta&T_1&0&.&.&.&.&.\\
.&.&0&T_1^\dagger&D_1&K_\mathrm{LM}^Q&0&.&.&.&.\\
.&.&.&0& K_\mathrm{ML}^Q& K_\mathrm{M} &\Theta_\mathrm{MR}&0&.&.&.\\
.&.&.&.&0&\Theta_\mathrm{RM}&D_2&\Theta&0&.&.\\
.&.&.&.&.&0&\Theta^\dagger&\Delta&\Theta&0&.\\
.&.&.&.&.&.&0&\Theta^\dagger&\Delta&\Theta&0\\
.&.&.&.&.&.&.&.&.&.&.\\
\end{array}\right)\;,
\label{dechamil}
\end{equation}
where $K_\mathrm{LM}^Q=H_\mathrm{LM}^Q-ES_\mathrm{LM}^Q$,
$K_\mathrm{ML}^Q=H_\mathrm{ML}^Q-ES_\mathrm{ML}^Q$ and
$K_\mathrm{M}=H_\mathrm{M}-ES_\mathrm{M}$.
The crucial point is that the new matrix ${\cal K}^{Q(\infty)}$ is still in the
desired tridiagonal form, but now the coupling matrices between principal layers
$\Theta$ are not singular. These are now $R\times R$ matrices obtained from the decimation
of the non-coupled degrees of freedom of the matrices $K_1^Q$ (the $C$ blocks).
Moreover the elimination of degrees of freedom achieved with the decimation scheme
is carried out only in the leads. The electronic structure of these is not updated
during the self-consistent procedure for evaluating the Green's function, and therefore
the information regarding the decimated degrees of freedom are not necessary. In contrast
the degrees of freedom of the scattering region are not affected by the decimation
or the rotation. Therefore the matrix $K_\mathrm{M}$ is unaffected by the decimation.

\begin{figure}[h]
\begin{center}
\includegraphics[width=7.5cm,clip=true]{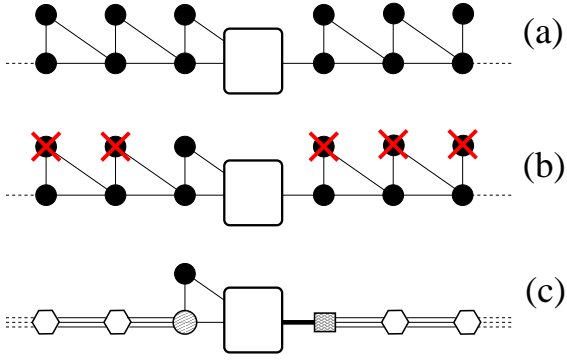}
\end{center}
\caption{Schematic representation of the decimation strategy for the rotated ${\cal K}$ 
matrix ${\cal K}^Q$.
Every symbol (dots, boxes ..) represents a collection of degrees of freedom (a matrix block)
and every line the coupling. (a) Original structure after the rotation ${\cal Q}$.
In the periodic leads the upper black dots represent the blocks $C$ of the matrix of equation
(\ref{matform}). The large white rectangular box represents the scattering region.
(b) The degrees of freedom marked with the red
crosses are decimated. (c) Final structure after decimation. The new white symbols represent
the leads degrees of freedom of the principal layers adjacent to the scattering region as
they appear after the decimation.}
\label{FigA2}
\end{figure}

In the decimated matrix ${{\cal K}^Q}^{(\infty)}$ new terms appear ($D_1$, $D_2$,
$T_1$ and $\Theta_\mathrm{MR}$). These arise from the specific structure of
the starting matrix ${\cal Q}^\dagger[{\cal H}-E{\cal S}]{\cal Q}$ and from the
fact that the complete system (leads plus scattering region) is not periodic.
In fact assuming that $j$ is the last principal layer of the left-hand side lead
and $l$ is the first layer of right-hand side lead, the decimation is carried out up to
$j-1$ to the left and starts from $l$ to the right of the scattering region. This
allows us to preserve the tridiagonal form of ${\cal K}$ and at the same time
to leave $K_\mathrm{M}$ unchanged. A schematic picture of the decimation strategy
is illustrated in figure \ref{FigA2}.

In practical terms all the blocks of the infinite matrix of equation (\ref{dechamil})
can be calculated by decimating auxiliary finite matrices. In particular

\begin{enumerate}
\item $\Delta$, $\Theta$ and $D_2$ are calculated by decimating both the $C$
matrices of the following finite $2N \times 2N$ matrix
\begin{equation}
\left(\begin{array}{cc}
\left(\begin{array}{cc}
C & B \\
B^\dagger & D
\end{array}\right) & \left[\begin{array}{cc}
0 & \bar{K}_1 \\
\end{array}\right] \\
\left[\begin{array}{c}
0 \\
{\bar{K}_1^{\dagger}}
\end{array}\right]  &  \left(\begin{array}{cc}
C & B \\
B^\dagger & D
\end{array}\right) \\
\end{array}\right)\longrightarrow
\left(\begin{array}{cc}
D_2 & \Theta \\
\Theta^\dagger & \Delta
\end{array}\right)\:,
\end{equation}
where $\bar{K}_1=\bar{H}_1-E\bar{S}_1$ and
\begin{equation}
\left(\begin{array}{cc}
C & B \\
B^\dagger & D
\end{array}\right)=H_0^Q-ES_0^Q\;.
\end{equation}

\item $D_1$ and $T_1$ are calculated by decimating only the upper $C$ matrix of the same
finite $2N \times 2N$ matrix
\begin{equation}
\left(\begin{array}{cc}
\left(\begin{array}{cc}
C & B \\
B^\dagger & D
\end{array}\right) & \left[\begin{array}{cc}
0 & \bar{K}_1 \\
\end{array}\right] \\
\left[\begin{array}{c}
0 \\
{\bar{K}^{\dagger}_1}
\end{array}\right]  &  \left(\begin{array}{cc}
C & B \\
B^\dagger & D
\end{array}\right) \\
\end{array}\right)\longrightarrow
\left(\begin{array}{cc}
D_2 & T_1 \\
T_1^\dagger & D_1
\end{array}\right)\:,
\end{equation}
where $D_1$ is $N\times N$, while $T_1$ is $R\times N$.

\item $\Theta_\mathrm{MR}$ is a $M\times R$ matrix obtained by decimating
the $C$ block of the following $(N+M)\times(N+M)$ matrix
\begin{equation}
\left(\begin{array}{cc}
0_\mathrm{M} & K_\mathrm{MR}^Q \\
K_\mathrm{RM}^Q  &  \left(\begin{array}{cc}
C & B \\
B^\dagger & D
\end{array}\right) \\
\end{array}\right)\longrightarrow
\left(\begin{array}{cc}
0_\mathrm{M} & \Theta_\mathrm{MR} \\
\Theta_\mathrm{RM} & D_2
\end{array}\right)\:,
\end{equation}
where $0_\mathrm{M}$ is the $M$-dimensional null matrix.

\end{enumerate}

Finally we are now in the position of calculating the self-energies. These are
obtained from the surface Green's functions for the rotated and decimated leads
(specified by the matrices $\Delta$ and $\Theta$) and have the following form
\begin{equation}
\Sigma_L=K_\mathrm{ML}^Q\left(-D_1-T_1^\dagger G_\mathrm{L}T_1\right)^{-1}K_\mathrm{LM}^Q
\end{equation}
and
\begin{equation}
\Sigma_R=\Theta_\mathrm{MR}\left[G_\mathrm{R}^{-1}-(D_2-\Delta)\right]^{-1}\Theta_\mathrm{RM}\:.
\end{equation}
Clearly our procedure not only regularizes the algorithm for calculating the self-energies, giving it
the necessary numerical stability, but also drastically reduces the degrees of freedom
(orbitals) needed for solving the transport problem. These go from $N$ (the dimension of the original
$H_1$ matrix) to $R$ (the rank of $H_1$). Usually $R\ll N$ and considerable computational overheads are
saved.

Finally it is important to note that usually the rank $R$ of $\left[\begin{array}{c} H_{1}\\ S_{1} \end{array}\right]$
is not necessarily the same of that of $\left[\begin{array}{c} H_{1}^\dagger\\ S_{1}^\dagger \end{array}\right]$ ($R^\prime$).
If $R^\prime<R$ the GSVD transformation must be performed over the matrices $H_{1}^\dagger$ and
$S_{1}^\dagger$. The procedure is similar to what described before but the final structure of the matrix
${\cal K}^Q$ is somehow different, and so should be the decimation scheme.

\section{Theoretical description}\label{AppB}

\subsection{Brief reminder of Keldish formalism}

The electronic part of the system we consider is described in a general way
by the following Hamiltonian \cite{Wag91}
\begin{equation}
H(\vec{r}_1,\vec{r}_2,t)=
H_0(\vec{r}_1,t)\delta(\vec{r}_1-\vec{r_2})+H_1(\vec{r}_1,\vec{r}_2)
\end{equation}
where $H_1$ accounts for the Coulomb interaction among electrons and $H_0$ stands
for all one-particle pieces of the Hamiltonian,
\begin{equation}
H_0(\vec{r},t)=H_\mathrm{kin}(\vec{r})+H_\mathrm{ei}(\vec{r})+
V_\mathrm{ext}(\vec{r},t)\;.
\end{equation}
$H_\mathrm{kin}$ and $H_\mathrm{ei}$ are respectively the kinetic energy and the Coulomb
interaction between electrons and nuclei, and $V_\mathrm{ext}$ is an external electrostatic
potentials applied to the system at $t=0$.

The electronic system evolves according to the time-independent Hamiltonian
$H(t_0=0^-)$ for all negative times, and uses $t_0$ as the synchronization time for all
pictures. Moreover we assume that  before $V_\mathrm{ext}$ is switched on
the system of interacting electrons is in thermodynamic
equilibrium at a chemical potential $\mu_0$.
We then prepare the density matrix $\rho$ at $t_0$ also. Therefore, expectation values of
observables
\begin{eqnarray}
\langle\hat{O}(t)\rangle&=& Tr\{\rho^S(t)\,\hat{O}^S(t)\}=
Tr\{\rho^H(t_0)\,\hat{O}^H(t)\}\nonumber\\
&=&\frac{1}{Z}\,Tr\{e^{-\beta\,(H(t_0)-\mu_0\,N)}\,
\hat{O}^H(t)\}
\end{eqnarray}
are described in terms of the density matrix of an interacting electron system at equilibrium.

These expectation values may be evaluated by using perturbation theory. Wick's theorem
may be applied only to ensembles of non-interacting electrons. In order to take advantage
of it, we must use a non-interacting density matrix such as
\begin{equation}
\rho_0(t_0)=\frac{e^{-\beta\,(H_0(t_0)-\mu_0\,N)}}{Z_0}
\end{equation}

\begin{figure}[h]
\begin{center}
\includegraphics[width=7.5cm,clip=true]{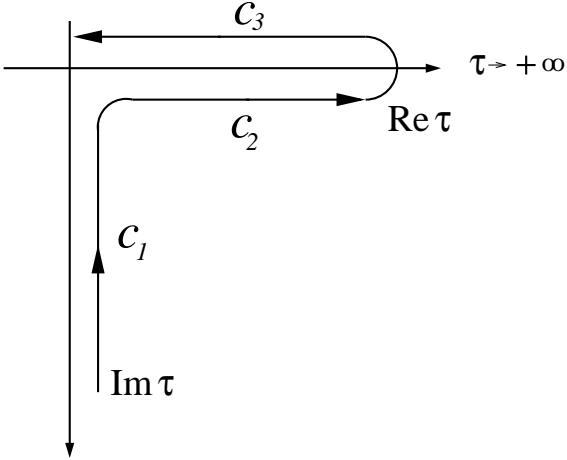}
\end{center}
\caption{Keldish contour $c$ in the complex $\tau$ plane.
The imaginary time path is called $c_1$ or $c_V$ in the text.
The segment lying below (above) the real-time axis is called $c_2$
($c_3$). The time loop $c_2$+$c_3$ is called $c_H$.}
\label{FigB1}
\end{figure}

We then define the Hamiltonian in the Schr\"odinger picture in the Keldish contour
of Fig. \ref{FigB1} of the complex $\tau$ plane as follows
\begin{equation}
K^S(\tau)= \left\{ \begin{array}{ll} H^S(\tau) &\tau\,\in\,c_H\\
H^S(0)-\mu_0\,N&\tau\,\in\,c_V
\end{array}\right.
\end{equation}
and analogous expressions for its noninteracting and interacting pieces, $K_0^S(\tau)$
and $K_1^S(\tau)$. We notice that the time variable is doubled valued along the real axis.
We must therefore distinguish whether any real time lies on the upper ($t^+$) or lower
($t^-$) branch.
%
%

We then define Heisenberg (H) and Dirac (D) pictures in the Keldish contour
whereby evolution of operators is provided by
\begin{eqnarray}
\hat{O}^H(t)&=&T_c\, \, e^{-i\,\int_{0^+}^{0^-} \,d\tau'\,K^S(\tau')}\,\hat{O}^S(t)=
T_c \,U(0^+,0^-)\,\hat{O}^S(t)
\nonumber\\
\hat{O}^D(t)&=&T_c\, \, e^{-i\,\int_{0^+}^{0^-} \,d\tau'\,K^S_0(\tau')}\,\hat{O}^S(t)=
T_c \,V_0(0^+,0^-)\,\hat{O}^S(t)\nonumber\\
&=&T_c\, \, e^{-i\,\int_{0^+}^{0^-} \,d\tau'\,K^D_1(\tau')}\,\hat{O}^H(t)=
T_c \,V_1(0^+,0^-)\,\hat{O}^H(t)\nonumber
\end{eqnarray}
where  $T_{c}$ is the  time ordering operator on $c$. Expectation values of operators
are then given by the statistical averages
\begin{equation}
\langle\hat{O}(t)\rangle=\frac{Z_0}{Z}\,Tr\{\rho_0\,V_1(-i\,\beta,0^+)\,\hat{O}(t)\}.
\end{equation}
which are evaluated in the non-interacting ensemble described by $\rho_0$.

The Green function of the system,
\begin{equation}
G(1,2) = \langle T_c\,\hat{\psi}(1)\,\hat{\psi}^\dagger(2) \rangle
\end{equation}
is a tool to compute the physical response of the system. Thus the electron charge and current
densities are simply
\begin{eqnarray}
\langle\hat{n}(1)\rangle&=& -i\,e\,G(1^+,1^-)\\
\langle\hat{\vec{\j}}(1)\rangle&=&-\frac{e\,\hbar}{2\,m}\,\left[\vec{\nabla}_1-\vec{\nabla}_2+
2\,i\,\vec{A}(1)\right]\,G(1^+,2)|_{2=1^-}\nonumber
\end{eqnarray}
and depend on the applied the external potential. Here, $\hat{\psi}\,,\hat{\psi}^\dagger$
denote creation and annihilation operators, and $(i)=(\vec{r}_i,t_i)$.

The Green's function can be proven to satisfy the following Dyson equation
\begin{eqnarray*}
\left[\,i\,\partial_{t_1} -H_0(1)\,\right]\,G(1,2)&=&\delta(1,2)+(\Sigma \otimes G)\,(1,2)
\nonumber\\
\left[\,-i\,\partial_{t_2} -H_0(2)\,\right]\,G(1,2)&=&\delta(1,2)+( G \otimes \Sigma)\,(1,2)
\end{eqnarray*}
where all time variables run through the Keldish contour $c$ and we follow the
conventional shorthand
\begin{equation}
(A\otimes B)(1,2)=\int_c \,dx_3\,dt_3 A(1,3)\,B(3,2)
\end{equation}

\subsection{Connection to TDDFT}
We now define a fictitious system of non-interacting electrons described by the
Hamiltonian \cite{Lee01}
\begin{equation}
H_s(\vec{r},t)=H_\mathrm{kin}(\vec{r})+V_s(\vec{r},t)
\end{equation}
where we assume that $H_s$ is constant and that the system is in thermodynamic
equilibrium at chemical potential $\mu_s$, all for negative times. We also assume
that $H_s$ may be exactly diagonalized. We do not split a perturbing piece from the
Hamiltonian and therefore, Heisenberg and Dirac pictures coincide. We again take $t_0$
as the synchronization and density-matrix preparation time. The density matrix
\begin{equation}
\rho_s(t_0)=\frac{e^{-\beta\,(H_s(t_0)-\mu_s\,N)}}{Z_0}
\end{equation}
and the time-evolution operator on the Keldish contour
\begin{equation}
U_s(0^+,0^-)=T_c\, \, e^{-i\,\int_{0^+}^{0^-} \,d\tau'\,K^S_s(\tau')}
\end{equation}
completely determine the expectation value of observables of this fictitious system
\begin{eqnarray}
\langle\hat{O}(t)\rangle_s= Tr\{\rho^S_s(t)\,\hat{O}^S(t)\}=
Tr\{\rho^H_s(t_0)\,\hat{O}^H(t)\}\nonumber
\end{eqnarray}

The Green's function $G_s(1,1')$, whose time variables also lie on $c$, satisfies
the equation of motion
\begin{eqnarray}
\left[\,i\,\partial_{t_1} -H_s(1)\,\right]\,G_s(1,2)&=&\delta(1,2)
\nonumber\\
\left[\,-i\,\partial_{t_2} -H_s(2)\,\right]\,G_s(1,2)&=&\delta(1,2)
\end{eqnarray}
and provides the physical response of the system
\begin{eqnarray}
n_s[V_s(1)]&=&  -i\,e\,G_s(1^+,1^-)\nonumber\\
\vec{\j}_s[V_s(1)]&=&
-\frac{e\,\hbar}{2\,m}\,\left[\vec{\nabla}_1-\vec{\nabla}_2+
2\,i\,\vec{A}(1)\right]\,G_s(1^+,2)|_{2=1^-}\nonumber
\end{eqnarray}

We define the action functionals
\begin{eqnarray}
R[V(x)]&=& i\, \ln \mathrm{Tr} \left[ \hat{U}(-i\,\beta,0)\right]\nonumber\\
R_s[V_s(x)]&=& i\, \ln \mathrm{Tr} \left[ \hat{U}_s(-i\,\beta,0)\right]
\end{eqnarray}
whose functional derivative with respect to the external potential provides an
alternative way to calculate the electronic density. We adjust the potential $V_s$
so that the densities of the two systems be equal,
\begin{equation}
\frac{\delta R[V(x)]}{\delta V(x)}=\frac{\delta R[V_s(x)]}{\delta V_s(x)}=\langle n(x)\rangle
\end{equation}
Notice that we do not require that the current densities of the
two systems be equal.

We now perform a Legendre transform to find the new actions
\begin{eqnarray}
S[n]&=&-R[V[n]]+\int dx\, n(x)\,V(x)\nonumber\\
S_s[n]&=&-R_s[V_s[n]]+\int dx\, n(x)\,V_s(x)
\end{eqnarray}
such that
\begin{eqnarray}
\frac{\delta S[n(x)]}{\delta n(x)}&=&V(x)\nonumber \\
\frac{\delta S_s[n(x)]}{\delta n(x)}&=&V_s(x)
\end{eqnarray}

We define an exchange-correlation functional
\begin{equation}
S_s[n]= S[n]+\frac{1}{2}\int\,dx\;[V_\mathrm{H}(x)\,n(x)+S_\mathrm{xc}[n]]
\end{equation}
whose functional derivative with respect to the density provides with a key
relationship between the external potential of the actual and fictitious systems,
\begin{eqnarray}
V_s(x)&=&V(x)+V_\mathrm{H}(x)+V_\mathrm{xc}(x)\nonumber\\
V_\mathrm{xc}[n(x)]&=&\frac{\delta S_\mathrm{xc}}{\delta n(x)}
\end{eqnarray}

Comparing the equations of motion of $G$ and $G_s$, we find an explicit
relationship between the Green's functions
\begin{equation}
G=G_s+G_s\otimes(\,\Sigma-V_\mathrm{H}-V_\mathrm{xc}\,)\otimes G
\end{equation}
and the Sham-Schl\"uter equation for the Self-energy
\begin{equation}
G_s\otimes (\,V_\mathrm{H}+V_\mathrm{xc}\,)\otimes G=G_s\otimes \Sigma[G]\otimes G
\end{equation}

Iterating these equations once means equating both Green's functions, $G=G_s$, which in
turn implies approximating $\vec{\j}$ by $\vec{\j}_s$.
The resulting integral equation for $\Sigma$ has
\begin{equation}
\Sigma[G](1,2)=(\,V_\mathrm{H}[n(1)]+V_\mathrm{xc}[n(1)]\,)\,\delta(1,2)
\end{equation}
as a trivial solution.
This simple approximation has the virtue that charge is conserved. This can be seen in two
alternative ways. First, the self-energy can be written as a $\Phi[G]$-derivable function.
Second, the fictitious system satisfies the continuity equation by construction.
Subsequent iterations improve the physical content of $G$, but we have used in our code
this lowest-order approximation.

The Green's function at two different times $G(t,t')$ can be viewed as the matrix
element $\langle t|G|t'\rangle$, sandwiched between two time states of the whole set $\{ |t\rangle \}$ of
times in the Keldish contour. We split the contour into the three pieces $c_1,\,c_2,\,c_3$
of Fig. 2, and define the corresponding three time subsets. The Green's function can then
be represented in matrix form as
\begin{eqnarray*}
\hat{G}_s(t,t')&=&
\left(\begin{array}{ccc}
G^{11}(t,t')&G^{12}(t,t')&G^{13}(t,t')\\
G^{21}(t,t')&G^{22}(t,t')&G^{23}(t,t')\\
G^{31}(t,t')&G^{32}(t,t')&G^{33}(t,t')
\end{array}
\right)\\&=&
\left(\begin{array}{ccc}
G^{c}(t,t')&G^<(t,t')&G^{13}(t,t')\\
G^>(t,t')&G^{ac}(t,t')&G^{13}(t,t')\\
G^{31}(t,t')&G^{31}(t,t')&G^{33}(t,t')
\end{array}
\right)
\end{eqnarray*}

The physical response of the system is therefore encapsulated in
the lesser Green function,
\begin{eqnarray}
n[V_s(1)]&=&  -i\,G_s^<(1^+,1^-)\nonumber\\
\vec{\j}[V_s(1)]&=&
-\frac{\hbar}{2\,m}\,\left[\vec{\nabla}_1-\vec{\nabla}_2+
2\,i\,\vec{A}(1)\right]\,G_s^<(1^+,2)|_{2=1^-}\nonumber
\end{eqnarray}

There are only five independent $G^{ij}$ out of the seven matrix elements displayed above.
A partial reduction to six matrix elements in the Green's function is achieved by the
non-unitary transformation
\begin{eqnarray*}
\check{G}_s(t,t')=L\,\tau_3\,\hat{G}_s(t,t')\,L^\dagger
=\left(\begin{array}{ccc}
G^\mathrm{R}&G^\mathrm{K}&\sqrt{2}\,G^{13}\\
0&G^\mathrm{A}&0\\0&\sqrt{2}\,G^{31}&G^33
\end{array}\right)
\end{eqnarray*}

Both matrices satisfy the matrix equations of motion
\begin{eqnarray}
\left[\,i\,\partial_{t_1} -H_s(1)\,\right]\, \hat{G}_s(1,2)&=&\delta(\vec{r}_1-\vec{r}_2)\,
\hat{\delta}(\tau_1-\tau_2)\nonumber\\
\left[\,i\,\partial_{t_1} -H_s(1)\,\right]\, \check{G}_s(1,2)&=&\delta(\vec{r}_1-\vec{r}_2)\,
\check{\delta}(\tau_1-\tau_2)\nonumber
\end{eqnarray}
where the time delta functions are defined as
\begin{equation}
\hat{\delta}\,,\check{\delta}=
\left(\begin{array}{ccc}
\delta(t_1-t_2)&0&0\\0&\pm\delta(t_1-t_2)&0\\0&0&\delta(\tau_1-\tau_2)
\end{array}\right)
\end{equation}
with the menos (plus) sign corresponding to the hat (check) delta function. We shall drop
the ``$s$'' subindex from now on, since all the discussion that follows is valid only for
the ficticious system of non-interacting electrons.

Physical response is customarily written in terms of $G^{\mathrm{R,A}}$ and $G^<$. Since there
is no linear transformation which allow to group them in one common matrix, one uses
Langreth's rules to find relationships among them.

\subsection{Equations of motion in the localized wave function basis}
The eigenstates of the system can be obtained by expanding them in the basis
of non-orthogonal states, $\Psi(n,\vec{r},t)=\sum_{i,\mu}\,c_{i\:\mu}(n,t)\,\psi_\mu(\vec{r}-\vec{R}_i)$, in terms of which
the Schr\"odinger equation reads
\begin{equation}
\left[i\,S_{i\mu\:j\nu}\,\partial_{t_1}-H_{i\mu\:j\nu}(t_1)\right]\,c_{j\:\nu}(n,t_1)=0
\end{equation}

Alternatively the equation of motion for either check or hat Green functions are
\begin{equation}
\left[\,i\,S_{i\mu\:k\lambda}\,\partial_{t_1} -H_{i\mu\:k\lambda}(t_1)\,\right]\,
G_{k\lambda\:j\nu}(t_1,t_2)=\delta_{ij}\,\delta_{\mu\nu}\,\hat{\delta}(t_1-t_2)\nonumber
\end{equation}

It is advantageous to perform a change of time variables from $t_1,t_2$ to
$T=1/2(t_1+t_2),t=t_1-t_2$. Then the Green's functions can be written as
\begin{equation}
G_{i\mu\:j\nu}(t_1,t_2)=G_{i\mu\:j\nu}(T,t)=\int\,\frac{dE}{2\,\pi}\,G_{i\mu\:j\nu}(T,E)\,e^{-i\,E\,t}
\nonumber
\end{equation}

The electron charge and current densities are found from
\begin{eqnarray}
n(\vec{r},T)&=&\sum_{i\mu\:j\nu}\,n_{i\mu\:j\nu}(\vec{r})\,\rho_{i\mu\:j\nu}(T)
\nonumber\\
\vec{\j}(\vec{r},T)&=&\sum_{i\mu\:j\nu}\,j_{i\mu\:j\nu}(\vec{r})\,\rho_{i\mu\:j\nu}(T)
\end{eqnarray}
where we have introduced the density matrix
\begin{equation}
\rho_{i\mu\:j\nu}(T)=\int\,\frac{dE}{2\,\pi\,i}\,G_{i\mu\:j\nu}^<(T,E)
\end{equation}
and
\begin{eqnarray}
&&n_{i\mu\:j\nu}(\vec{r}_1,\vec{r}_2)=e\,\psi_{\mu}(\vec{r}_1-\vec{R}_i)\,\,\psi_\nu(\vec{r}_2-\vec{R}_j)
\nonumber\\
&&\vec{\j}_{i\mu\:j\nu}(\vec{r})=-\frac{i\,e\,\hbar}{2\,m}\left(\vec{\nabla}_{\vec{r}_1}-
\vec{\nabla}_{\vec{r}_2}+2i\,\vec{A}(\vec{r}_1)\right)\,n_{i\mu\:j\nu}(\vec{r}_1,\vec{r}_2)|_{r_2=r_1}
\nonumber
\end{eqnarray}

The electric current through a given surface $S$ is obtained by integrating the current density
over such surface
\begin{eqnarray}
I&=&\int_S d\vec{S}\cdot\vec{\j}(\vec{r},t_1)=\sum_{i\mu\:j\nu}\rho_{i\mu\:j\nu}(t_1)
\int_S\,d\vec{S}\cdot\vec{\j}_{i\mu\:j\nu}\nonumber\\
&=&\sum_{i'\mu'\:j'\nu'}\rho_{i'\mu'\:j'\nu'}\,H_{i'\mu'\:j'\nu'}
\end{eqnarray}
where only those bonds $(i'\mu\:j'\nu')$ pierced by the surface contribute to the summation.

If the system is in thermodynamic equilibrium, the population of electrons does not
depend on time, and follows the Fermi-Dirac distribution function $f(\epsilon)$. Thereby
the lesser Green function can be written in terms of the retarded one as
\begin{eqnarray}
G^<_{i\mu\:j\nu}(E)&=&f(E)\,\left[G^\mathrm{A}_{i\mu\:j\nu}(E)-G^\mathrm{R}_{i\mu\:j\nu}(E)\right]=\nonumber\\
&=&-2\,i\,f(E)\,\mathrm{Im}\,\left[G^\mathrm{R}_{i\mu\:j\nu}(E)\right]=\nonumber\\
&=&2\,\pi\,i\,f(E)\,\sum_n\,c_{i\mu}(n)\,c_{j\nu}^*(n)\,\delta(E-\epsilon_n)\nonumber
\end{eqnarray}
where $\epsilon_n$ are the eigenvalues of the Hamiltonian. Therefore, the density matrix can
be obtained from the wave function coefficients by just diagonalising the Hamiltonian,
\begin{eqnarray}
\rho_{i\mu\:j\nu}&=&-\frac{1}{\pi}\int{\mathrm d}E\,f(E)\,\mathrm{Im}\,\left[G^\mathrm{R}_{i\mu\:j\nu}(E)\right]=\nonumber\\
&=&\sum_n\,c_{i\mu}(n)\,c_{j\nu}^*(n)\,f(\epsilon_n)
\end{eqnarray}

\subsection{The extended molecule setup}
We now wish to partition the Green functions according to the system
setup of Fig. 1, where the left and right leads remain in thermodynamic equilibrium defined
by $\mu_{\mathrm{L/R}}=E_\mathrm{F}\pm e\,V/2$ at all times. The extended molecule is also in thermodynamic
equilibrium for times $t<0$. The Hamiltonian for negative times
\begin{equation}
{\cal H}(t<0)=h(t)=
\left(\begin{array}{ccc}
{\cal H}_\mathrm{L}+eV/2\,{\cal S}_\mathrm{L}&0&0\\
0&H_\mathrm{M}&0\\
0&0&{\cal H}_\mathrm{R}-eV/2\:{\cal S}_\mathrm{R}\\
\end{array}\right)\:,\nonumber
\end{equation}
serves to define the reference equilibrium hat and check Green functions, that
satisfy the equations of motion
\begin{equation}
\left[\,i\,S_{i\mu,k\lambda}\,\partial_{t_1}
-h_{i\mu,k\lambda}(t_1)\,\right]\,
G_{k\lambda,j\nu}(t_1,t_2)=\delta_{i,j}\,\delta_{\mu,\nu}\,\hat{\delta}(t_1-t_2)\nonumber
\end{equation}
For instance, the equation of motion for the retarded Green function in
frequency domain is just
\begin{widetext}
\begin{equation}
\left(\begin{array}{ccc}
(\epsilon^+-eV/2){\cal S}_\mathrm{L}-{\cal H}_\mathrm{L}&0&0\\
0&\epsilon^+S_\mathrm{M}-H_\mathrm{M}&0\\
0&0&(\epsilon^++eV/2){\cal S}_\mathrm{RM}-{\cal H}_\mathrm{RM}\\
\end{array}\right)
\left(\begin{array}{ccc}
{\cal G}_\mathrm{L}^{0\mathrm{R}}&0&0\\
0&G_\mathrm{M}^{0\mathrm{R}}&0\\
0&0&{\cal G}_\mathrm{R}^{0\mathrm{R}}\\
\end{array}\right)=
\left(\begin{array}{ccc}
{\cal I}&0&0\\
0&I_\mathrm{M}&0\\
0&0&{\cal I}\\
\end{array}\right)\;,
\end{equation}
while the lesser Green function is
\begin{equation}
{\cal G}^{0<}(E)=\left(\begin{array}{ccc}
\left({\cal G}_\mathrm{L}^{0\mathrm{A}}(E)-{\cal G}_\mathrm{L}^{0\mathrm{R}}(E)\right)\,f(E-eV/2)&0&0\\0&G_M^{0<}(E)&0\\0&0&
\left({\cal G}_\mathrm{R}^{0\mathrm{A}}(E)-{\cal G}_\mathrm{R}^{0\mathrm{R}}(E)\right)\,f(E+eV/2)
\end{array}\right)
\end{equation}
\end{widetext}

The extended molecule is contacted by the electrodes at time $t=0$, through the
Hamiltonian matrix elements
\begin{equation}
V_{ext}=
\left(\begin{array}{ccc}
0&{\cal H}_\mathrm{LM}&0\\
{\cal H}_\mathrm{ML}&0&{\cal H}_\mathrm{MR}\\
0&{\cal H}_\mathrm{RM}&0\\
\end{array}\right)\:F(t),
\end{equation}
where $F(t)$ is zero for negative times, and 1 for times larger
than a certain characteristic time $\tau_{M}$. The perturbation
$V_{ext}$ drives the core of the extended molecule out of
equilibrium for positive times, by populating it with a
distribution of electrons that does not follow Fermi-Dirac
statistics. The distribution function of the non-contacted
molecule $g_M^<(E)$ is completely washed out, but the
density-matrix of the system can still be determined from the
equations of motion of the lesser and retarded Green functions.

We seek to solve the equations of motion for times $t\gg\tau_{M}$ where all transient effects
have vanished. This means, first, that $G^{1,3}=G^{3,1}=0$, so that the matrix hat and check
Green functions are block-diagonal; second, that the Hamiltonian is simply ${\cal H}$; and third,
that Green functions do not depend on the time variable T.

The retarded Green function is, simply,
\begin{widetext}
\begin{equation}
\left(\begin{array}{ccc}
(\epsilon^+-eV/2)\,{\cal S}_\mathrm{L}-{\cal H}_\mathrm{L}&\epsilon^+{\cal S}_\mathrm{LM}-{\cal H}_\mathrm{LM}&0\\
\epsilon^+{\cal S}_\mathrm{ML}-{\cal H}_\mathrm{ML}&\epsilon^+S_\mathrm{M}-H_\mathrm{M}&\epsilon^+{\cal S}_\mathrm{MR}-{\cal H}_\mathrm{MR}\\
0& \epsilon^+{\cal S}_\mathrm{RM}-{\cal H}_\mathrm{RM}&(\epsilon^++eV/2)\,{\cal S}_\mathrm{R}-{\cal H}_\mathrm{R}\\
\end{array}\right)
\left(\begin{array}{ccc}
{\cal G}_\mathrm{L}^{\mathrm R}&{\cal G}_\mathrm{LM}^{\mathrm R}&{\cal G}_\mathrm{LR}^{\mathrm R}\\
{\cal G}_\mathrm{ML}^{\mathrm R}&G_\mathrm{M}^{\mathrm R}&{\cal G}_\mathrm{MR}^{\mathrm R}\\
{\cal G}_\mathrm{RL}^{\mathrm R}&{\cal G}_\mathrm{RM}^{\mathrm R}&{\cal G}_\mathrm{R}^{\mathrm R}\\
\end{array}\right)=
\left(\begin{array}{ccc}
{\cal I}&0&0\\
0&I_\mathrm{M}&0\\
0&0&{\cal I}\\
\end{array}\right)\;,
\label{EqB3}
\end{equation}
\end{widetext}

The lesser Green function is better expressed in terms of the retarded and advanced,
 and the reference equilibrium lesser Green function, by using Langreth's rules, as
 \begin{equation}
 {\cal G}^<={\cal G}^\mathrm{R}\,({\cal G}^{0\mathrm{R}})^{-1}\,{\cal G}^{0<}\,({\cal G}^{0\mathrm{A}})^{-1}\,{\cal G}^\mathrm{A}
 \end{equation}
Straightforward matrix algebra then leads to Eqs. (\ref{density}) and (\ref{current}).

%

\begin{thebibliography}{99}

\bibitem{STM}
G. Binnig, H. Rohrer, Ch. Gerber and E. Weibel, Phys. Rev. Lett. {\bf 50}, 120 (1983).

\bibitem{memory}
P.J.~Kuekes, J.R.~Heath, and R.S.~Williams,
US Patent, number 6128214 (Hewlett-Packard), October 2000.

\bibitem{gates1}
C.P.~Collier, E.W.~Wong, M.~Belohradsky, F.M.~Raymo, J.F.~Stoddart, P.J.~Kuekes,
R.S.~Williams and J.R.~Heath,
Science {\bf 285}, 391 (1999).

\bibitem{gates2}
Y.~Huang, X.~Duan, Y.~Cui, L.J.~Lauhon, K.-H.~Kim, C.M.~Lieber,
Science {\bf 294}, 1313 (2001).

\bibitem{PC1}
N. Garc\'{\i}a, M. Mu\~noz, and Y.-W. Zhao, Phys. Rev. Lett. {\bf
82}, 2923 (1999).

\bibitem{PC2}
J.J. Versluijs, M.A. Bari, and J.M.D. Coey, Phys. Rev. Lett. {\bf 87}, 026601 (2001).

\bibitem{NO2}
P.~Qi, O.~Vermesh, M.~Grecu, A.~Javey, Q.~Wang, H.~Dai, S.~Peng and K.~Cho,
Nano Lett. {\bf 3}, 347 (2003).

\bibitem{nerve}
J.P.~Novak, E.S.~Show, E.J.~Houser, D.~Park, J.L.~Stepnowski, and R.A.~McGill,
Appl. Phys. Lett. {\bf 83}, 4026 (2003).

\bibitem{virus}
F.~Patolsky, et al.,
Proc. Natl. Acad. Sci. USA {\bf 101}, 14017 (2004).

\bibitem{chem}
Y.~Cui, W.~Qingqiao, P.~Hongkun, and C.M.~Lieber,
Science {\bf 293}, 1289 (2001).

\bibitem{med}
J.P.~Heath, M.E.~Phelps and L.~Hood,
Mol. Im. Biol. {\bf 5}, 312 (2003).

\bibitem{GF1}
S.~Datta, {\em Electronic Transport in Mesoscopic Systems}, Cambridge University Press,
Cambridge, UK, 1995.

\bibitem{GF2}
H.~Haug and A.~P. Jauho, {\em Quantum Kinetics in Transport and Optics of Semiconductors},
Springer, Berlin, 1996.

\bibitem{GF3}
C.~Caroli, R.~Combescot, P.~Nozieres, and D.~Saint-James,
J. Phys. C: Solid State Phys. {\bf 5}, 21 (1972).

\bibitem{GF4}
J.~Ferrer, A.~Mart\'{\i}n-Rodero, and F.~Flores,
Phys. Rev. B {\bf 38}, R10113 (1988).

\bibitem{WF1}
N.~D. Lang,
Phys. Rev. B {\bf 36}, R8173 (1987).

\bibitem{WF2}
N.~D. Lang, Phys. Rev. B {\bf 52}, 5335 (1995).

\bibitem{WF3}
M.~Di Ventra, S.~T. Pantelides, and N.~D. Lang,
Phys. Rev. Lett. {\bf 84}, 979 (2000).

\bibitem{ME1}
C.W.J. Beenakker,
Phys. Rev. B {\bf 44}, 1646 (1991)

\bibitem{ME2}
R.~Gebauer and R.~Car,
Phys. Rev. B {\bf 70}, 125324 (2004).

\bibitem{ME3}
B. Muralidharan, A.W. Ghosh, S.K. Pati and S. Datta,
cond-mat/0505375

\bibitem{StefRev}
S. Sanvito, in {\it Handbook of Computational Nanotechnology},
American Scientific Publishers (Stevenson Ranch, California, 2005),
also cond-mat/0503445.

\bibitem{stef1}
G. Stefanucci and C.-O. Almbladh,
Phys. Rev. B {\bf 69}, 195318 (2004).

\bibitem{DFT}
H.~Hohenberg and W.~Kohn,
Phys. Rev. {\bf 136}, B864 (1964).

\bibitem{KohnSham}
W.~Kohn and L.J.~Sham,
Phys. Rev. {\bf 140}, A1133 (1965).

\bibitem{NEGF1}
J.~Taylor, H.~Guo, and J.~Wang,
Phys. Rev. B {\bf 63}, 245407 (2001).

\bibitem{NEGF2}
Y.~Xue, S.~Datta and M.A.~Ratner,
Chem. Phys. {\bf 281}, 151 (2002).

\bibitem{NEGF3}
M.~Brandbyge, J.-L. Mozos, P.~Ordej\'on, J.~Taylor, and K.~Stokbro, 
Phys. Rev. B {\bf 65}, 165401 (2002).

\bibitem{NEGF4}
J.~J. Palacios, A.~J. P\'erez-Jim\'enez, E.~Louis, E.~SanFabi\'an, and J.~A.
  Verg\'es, Phys. Rev. B {\bf 66}, 035322 (2002).

\bibitem{NEGF5}
A.~Pecchia and A.~Di Carlo,
Rep. Prog. Phys. {\bf 67}, 1497 (2004).

\bibitem{cini}
M. Cini, Phys. Rev. B {\bf 22}, 5887 (1980).

\bibitem{Gross}
S. Kurth, G. Stefanucci, C.-O. Almbladh, A. Rubio, and E.K.U. Gross,
Phys. Rev. B {\bf 72}, 035308 (2005).

\bibitem{TDDFT}
E. Runge, and E.K.U. Gross, Phys. Rev. Lett. {\bf 52}, 997 (1984).

\bibitem{Tchav}
A.P. Horsfield, D.R. Bowler and A.J. Fisher, T.N. Todorov and C.G. Sanchez,
J. Phys. Condens. Matter {\bf 16}, 8251 (2004).

\bibitem{Max1}
N. Bushong, N. Sai, M. di Ventra,
cond-mat/0504538

\bibitem{Max2}
N. Sai, M. Zwolak, G. Vignale and M. di Ventra,
cond-mat/0411098.

\bibitem{Kieron1}
K. Burke, M. Koentopp and F. Evers, cond-mat/0502385.

\bibitem{Smeagol1}
A.R. Rocha and V.M. Garc\'{\i}a-Su\'arez and S.W. Bailey and C.J.
Lambert and J. Ferrer and S. Sanvito, {\it Smeagol}: Spin and
Molecular Electronics in Atomically Generated Orbital Landscapes.
http://www.smeagol.tcd.ie/

\bibitem{rgf}S.~Sanvito, C.~J.~Lambert, J.~H.~Jefferson, and A.M.~Bratkovsky,
Phys. Rev. B {\bf 59}, 11936-11948 (1999).

\bibitem{Smeagol2}
A.R. Rocha and V.M. Garc\'{\i}a-Su\'arez and S.W. Bailey and C.J.
Lambert and J. Ferrer and S. Sanvito, Nature Materials {\bf 4},
335 (2005).

\bibitem{MZ}
O.F.~Sankey and D.J.~Niklewski, Phys. Rev. B {\bf 40}, 3979
(1989).

\bibitem{KEL1}L.P. Kadanoff, and G. Baym, {\em Quantum Statistical Mechanics},
W.A. Benjamin, Menlo Park, CA, 1962.

\bibitem{KEL2}L.V. Keldysh, Sov. Phys. JETP {\bf 20}, 1018 (1965).

\bibitem{meir}Y. Meir and N. S. Wingreen, Phys. Rev. Lett. {\bf
68}, 2512 (1992).

\bibitem{LBYP}M.~B\"uttiker, Y.~Imry, R.~Landauer and S.~Pinhas, Phys. Rev. B {\bf 31},
6207 (1985).

\bibitem{fisherlee}D.S.~Fisher and P.A.~Lee, Phys. Rev. B {\bf 23}, R6851 (1981).

\bibitem{TCH1}Some caution should be taken in selecting the plane for evaluating the conductance
when the basis set is not complete as in the case of LAO basis sets. See for instance T. Todorov, J.~Phys.:
Condens. Matter {\bf 13}, 10125 (2001); T. Todorov, J.~Phys.:
Condens. Matter {\bf 14}, 3049 (2002).

\bibitem{lang}
A.~R.Williams, P.~J. Feibelman and N.~D. Lang, Phys. Rev. B {\bf 26},
5433 (1982).

\bibitem{nardelli}M.B. Nardelli, Phys. Rev. B {\bf 60}, 7828 (1999).

\bibitem{Alex}
A.~R. Rocha and S.~Sanvito.
Phys. Rev. B {\bf 70}, 094406 (2004).

\bibitem{smeagolsic}
C.~Toher, A.~Filippetti, S.~Sanvito and K.~Burke,
to appear in Phys. Rev. Lett., also cond-mat/0506244.

\bibitem{siesta}J.~M.~Soler, E.~Artacho, J.~D.~Gale, A.~Garc\'{\i}a, J.~Junquera, P.~Ordej\'on and
D.~Sanchez-Portal, J. Phys. Cond. Matter {\bf 14}, 2745-2779 (2002).

\bibitem{KB} L.~Kleinman, D.M.~Bylander, Phys. Rev. Lett. {\bf 48}, 1425 (1982)

\bibitem{SIESTAbasis1} J.~Junquera, O.~Paz, D.~S\'anchez-Portal and E.~Artacho,
{Phys.~Rev.~B} {\bf 64}, 235111 (2001)

\bibitem{SIESTAbasis2}E.~Anglada, J.~M.~Soler, J.~Junquera and E.~Artacho,
{Phys.~Rev.~B} {\bf 66}, 205101 (2002)

\bibitem{Pettifor}D.~Pettifor, {\em Bonding and structure of molecules and solids},
Oxford University Press, Oxford, 2002.

\bibitem{ruitenbeek}N.~Agra{\"i}t, A.~Levy Yeyati and J.M.~van~Ruitenbeek, Phys. Rep. {\bf 377},
81 (2003).

\bibitem{Rodriguez}V.~Rodrigues, J.~Bettini, A.R.~Rocha, L.G.C.~Rego and D.~Ugarte,
Phys.~Rev.~B {\bf 65}, 1534024 (2002)

\bibitem{taka}H.~Ohnishi, Y.~Kondo, K.~Takayanagi,
Nature (London) {\bf 395}, 780 (1998).

\bibitem{ugarte}P.Z.~Coura, S.B.~Legoas, A.S.~Moreira, F.~Sato,
V.~Rodrigues, S.O.~Dantas, D.~Ugarte, and D.S.~Galv\v{a}o,
Nano Lett. {\bf 4}, 1187 (2004).

\bibitem{landman}U.~Landman, W.D.~Luedtke, B.E.~Salisbury, and R.L. Whetten,
Phys.~Rev.~Lett. {\bf 77}, 1362 (1996).

\bibitem{rego_rocha}L.G.C.~Rego, A.R.~Rocha, V.~Rodrigues, and D.~Ugarte,
Phys.~Rev.~B {\bf 67}, 045412 (2003).

\bibitem{RodriguezLett}V.~Rodrigues, T.~Fuhrer, and D.~Ugarte,
Phys.~Rev.~Lett. {\bf 85}, 4124 (2000).

\bibitem{vic1}V.M.~Garc{\'{\i}}a-Su\'arez, A.R.~Rocha, S.W.~Bailey, C.J.~Lambert, S.~Sanvito,
and J.~Ferrer, Phys. Rev. B {\bf 72}, 045437 (2005).

\bibitem{vic2}V.M.~Garc{\'{\i}}a-Su\'arez, A.R.~Rocha, S.W.~Bailey, C.J.~Lambert, S.~Sanvito,
and J.~Ferrer, cond-mat/0505487.

\bibitem{GMR1}M.N.~Baibich, J.M.~Broto, A.~Fert, F.~Nguyen Van Dau, F.~Petroff,
P.~Etienne, G.~Creuzet, A.~Friederich and J.~Chazelas,
Phys.~Rev.~Lett. {\bf 61}, 2472 (1988).

\bibitem{GMR2}G.~Binasch, P.~Gr\"unberg, F.~Saurenbach and W.~Zinn,
Phys. Rev. B {\bf 39}, 4828 (1989).

\bibitem{bruce} K.~Tsukagoshi, B.~W. Alphenaar, and H.~Ago, Nature (London) {\bf 401}, 572-574 (1999).

\bibitem{Aws1} M.~Ouyang and D.~D. Awschalom, Science {\bf 301}, 1074-1078 (2003).

\bibitem{Shi} Z.~H.~Xiong, D.~Wu, Z.~Valy~Vardeny and J.~Shi,
Nature (London) {\bf 427}, 821-824 (2004).

\bibitem{Dediu} V.~Dediu, M.~Murgia, F.C.~Matacotta, C.~Taliani, and S.~Barbanera,
Solid State Commun. {\bf 122}, 181-184 (2002).

\bibitem{Ralph} J.~R.~Petta, S.~K.~Slater and D.~C.~Ralph,
Phys. Rev. Lett. {\bf 93}, 136601 (2004).

\bibitem{bruno}P.~Bruno, Phys. Rev. Lett. {\bf 83}, 2425 (1999).

\bibitem{viret1}M.~Viret, S.~Berger, M.~Gabureac, F.~Ott, D.~Olligs, I.~Petej,
J.F.~Gregg, C.~Fermon, G.~Francinet and G.~Le~Goff, Phys. Rev. B {\bf 66}, 220401(R)
(2002).

\bibitem{garcia2001}N.~Garc\'{\i}a, M.~Mu\~noz, G.~G.~Qian, H.~Rohrer, I.~G.~Saveliev and
Y.~W.~Zhao, Appl. Phys. Lett. {\bf79}, 4550 (2001).

\bibitem{zchopra}S.~Z.~Hua and H.~D.~Chopra, Phys. Rev. B {\bf 67}, 060401(R)
(2003).

\bibitem{oscar3}O.~C{\'e}spedes, A.~R. Rocha~S. Lioret, M.~Viret, C.~Dennis, J.~F. Gregg,
S.~van Dijken, S.~Sanvito and J.~M.~D. Coey, J. Magn. Magn. Matter {\bf 272-276}, 1571 (2004).

\bibitem{viret2}M.~Gabureac, M.~Viret, F.~Ott and C.~Fermon, Phys. Rev. B {\bf 69},
100401(R) (2004)

\bibitem{Bahn}S.R.~Bahn and K.W.~Jacobsen, Phys. Rev. Lett. {\bf 87}, 266101 (2001).

\bibitem{Maria}M. Stamenova, S.~Sanvito and T.~Todorov, cond-mat/0505134.

\bibitem{mertig}A.~Bagrets, N.~Papanikolaou and I.~Mertig, Phys. Rev. B {\bf 70}, 064410 (2004).

\bibitem{palacios}D.~Jacob, J.~Fern\'andez-Rossier and J.J.~Palacios,
Phys. Rev. B {\bf 71}, 220403(R) (2004).

\bibitem{smit} R. H. M. Smit, Y. Noat, C. Untiedt, N. D. Lang,
M. C. van Hemert, J. M. van Ruitenbeek, Nature (London) {\bf 419},
906 (2002).

\bibitem{cuevas} J. C. Cuevas, J. Heurich, F. Pauly, W. Wenzel,
and G. Sch\"on, Nanotechnology {\bf 14}, R29 (2003).

\bibitem{garcia-pala} Y. Garc\'{\i}a, J. J. Palacios, E. SanFabi\'an,
J. A. Verg\'es, A. J. P\'erez-Jim\'enez, and E. Louis, Phys. Rev.
B {\bf 69}, 041402(R) (2004).

\bibitem{thygesen} D. Djukic, K.S. Thygesen, C. Untiedt, R.H. M. Smit, K. W. Jacobsen
and J. M. van Ruitenbeek, arXiv:cond-mat/0409640; K. S. Thygesen
and K. W. Jacobsen, arXiv:cond-mat/0411088.




\bibitem{lapack}E.~Anderson, Z.~Bai, C.~Bischof, S.~Blackford, J.~Demmel, J.~Dongarra,
J.~Du Croz, A.~Greenbaum, S.~Hammarling, A.~McKenney, and D.~Sorensen,
{\em {LAPACK} Users' Guide},  Society for Industrial and Applied Mathematics,
Philadelphia, PA (1999).

\bibitem{Wag91} M. Wagner, Phys. Rev. B {\bf 44}, 6104 (1991).

\bibitem{Lee01} R. van Leeuwen, Int. J. Mod. Phys. B {\bf 15}, 1969
(2001).




%
%
%
%
%
%
%
%
%
%
%
%
%
%
%
%
%
%
%
%
%
\end{thebibliography}

\end{document}